\documentclass[10pt,journal,compsoc]{IEEEtran}

\usepackage{ifpdf}
\usepackage{cite}
\usepackage{amsmath,amssymb,amsfonts}
\usepackage{algorithmic}
\usepackage{graphicx, adjustbox}
\usepackage{array}
\usepackage{enumitem}
\usepackage{breqn}
\usepackage{ dsfont }
\usepackage{ upgreek }
\usepackage{textcomp}
\usepackage{multirow}
\usepackage{algorithm}
\usepackage{algorithmic}
\usepackage{caption}
\usepackage{pgfplots}
\usepackage{tkz-euclide}
\usepackage{subcaption}
\usepackage{tikz}{\tiny }
\usepackage{hyperref}
\usepackage{balance}

\DeclareMathOperator*{\maximize}{maximize}
\usepackage{fancyhdr}

\fancypagestyle{firstpage}
{
	\fancyhead[L]{Accepted at IEEE Transactions on Mobile Computing}    
	\fancyhead[R]{ }
}

\begin{document}
	
	\title{Cooperative Task Offloading and Block Mining in Blockchain-based Edge Computing with Multi-agent Deep Reinforcement Learning}
	
	\author{Dinh C. Nguyen,~\IEEEmembership{Member,~IEEE,} 
		Ming Ding,~\IEEEmembership{Senior Member,~IEEE,} Pubudu N. Pathirana,~\IEEEmembership{Senior Member,~IEEE,} 
		Aruna Seneviratne,~\IEEEmembership{Senior Member,~IEEE}, Jun Li,~\IEEEmembership{Senior Member,~IEEE}, \\ and H. Vincent Poor,~\IEEEmembership{Fellow,~IEEE}
		
		\IEEEcompsocitemizethanks{
			\IEEEcompsocthanksitem Dinh C. Nguyen and Pubudu N. Pathirana are with School of Engineering, Deakin University, Waurn Ponds, VIC 3216, Australia (e-mails: \{cdnguyen, pubudu.pathirana\}@deakin.edu.au).
			\IEEEcompsocthanksitem Ming Ding is with the Data61, CSIRO, Australia (email: ming.ding@data61.csiro.au).
			\IEEEcompsocthanksitem Aruna Seneviratne is with School of Electrical Engineering and Telecommunications, University of New South Wales (UNSW), NSW, Australia (e-mail: a.seneviratne@unsw.edu.au).
			\IEEEcompsocthanksitem Jun Li is with School of Electrical and Optical Engineering, Nanjing University of Science and Technology, Nanjing 210094, China (e-mail: jun.li @njust.edu.cn).
			\IEEEcompsocthanksitem H. Vincent Poor is with the Department of Electrical Engineering, Princeton University, Princeton, NJ 08544 USA (e-mail: poor@princeton.edu).
			\IEEEcompsocthanksitem This work was supported in part by the CSIRO Data61, Australia, and in part by U.S. National Science Foundation under Grant CCF-1908308. The work of Jun Li was supported by National Natural Science Foundation of China under Grant 61872184.
	}}
	

	\IEEEtitleabstractindextext{%
		\begin{abstract}
			The convergence of mobile edge computing (MEC) and blockchain is transforming the current computing services in mobile networks, by offering task offloading solutions with security enhancement empowered by blockchain mining. Nevertheless, these important enabling technologies have been studied separately in most existing works. This article proposes a novel cooperative task offloading and block mining (TOBM) scheme for a blockchain-based MEC system where each edge device not only handles data tasks but also deals with block mining for improving the system utility. To address the latency issues caused by the blockchain operation in MEC, we develop a new Proof-of-Reputation consensus mechanism based on a lightweight block verification strategy. A multi-objective function is then formulated to maximize the system utility of the blockchain-based MEC system, by jointly optimizing offloading decision, channel selection, transmit power allocation, and computational resource allocation. We propose a novel distributed deep reinforcement learning-based approach by using a multi-agent deep deterministic policy gradient algorithm. We then develop a game-theoretic solution to model the offloading and mining competition among edge devices as a potential game, and prove the existence of a pure Nash equilibrium. Simulation results demonstrate the significant system utility improvements of our proposed scheme over baseline approaches. 
		\end{abstract}
		
		
		}
	

	\maketitle
	\IEEEdisplaynontitleabstractindextext
	\IEEEpeerreviewmaketitle
	
	\thispagestyle{firstpage}
	\section{Introduction}
	Recent advances in Internet of Things (IoT) have promoted the proliferation of numerous mobile applications that mostly rely on edge devices (EDs), e.g., laptops, tablets, and smartphones, to collect data from IoT sensors to serve end users. To meet the increasing users' computation demand, mobile edge computing (MEC) has been proposed as a promising technique to improve the computation experience of EDs, by offloading computationally intensive tasks to a nearby MEC server located at a base station (BS) \cite{add13}. Mapping each offloading process to a specific application, multiple distributed EDs naturally share computation and communication resources of the BS to handle data tasks without device's battery depletion. Task offloading with MEC thus becomes  a viable solution to satisfy various EDs' computation demands and  enhances the quality of experience (QoE) of end users. 
	
	However, the design of an efficient task offloading scheme for MEC systems still faces non-trivial challenges. Each ED always aims to maximize its individual utility by occupying as many edge resources (e.g., channel spectrum, CPU frequency) as possible, which is likely to cause network traffic congestion and user interference. \textcolor{black}{The heterogeneous resource requirements of multiple IoT data tasks, e.g., different resource allocations needed for handling different data tasks, and the heterogeneous features of real-time IoT data tasks, e.g., computation deadlines and data task sizes, pose challenges for the design of offloading strategies for all EDs. Moreover, the lack of prior information on system statistics in practical multi-user MEC systems, e.g., channel state and edge computational resource state, makes it challenging to derive an optimal offloading solution for each ED.} Therefore, it is of the utmost importance to develop a intelligent and self-organized offloading scheme to guide the offloading actions of all EDs in the distributed MEC systems. 
	
	Furthermore, the dynamic communications between IoT devices, EDs, and the MEC server and the migration of IoT data tasks across the MEC network potentially cause security vulnerabilities.  \textcolor{black}{Recent works \cite{add13,add11,add12} have mainly focused on computation offloading designs for task scheduling and resource allocation, with the lack of considering security aspects in MEC networks. } Fortunately, blockchain has been envisioned as a strong candidate to enhance security of MEC systems \cite{consensus2}. In fact, blockchain is able to provide high degrees of security and trust for MEC  by employing the community verification among edge nodes via mining mechanisms such as Delegated Proof of Stake (DPoS) \cite{consensus2} without requiring any central authority. 
	
	\textcolor{black}{In this context, the use of blockchain is highly desirable to support edge computing systems \textcolor{black}{\cite{xiong2018mobile, guo2019blockchain,kangblockchain2019}}. Specifically, blockchain decentralizes the MEC system where edge nodes can communicate with  each other via the peer-to-peer network over the decentralized data ledger. Different from traditional MEC systems that often rely on a central server to coordinate the MEC operation, blockchain helps build decentralized edge communications without the need for a single authority, which eliminates the risks of single-point failure. This feature is very useful in practical application scenarios, e.g., decentralized edge data sharing and decentralized edge data caching in MEC networks. Another motivation behind the integration of blockchain in MEC is its immutability that makes edge data records, e.g., IoT data, unchangeable once they are stored on the ledger \cite{add7}. By deploying immutable transaction ledgers, EDs can establish reliable communications to perform heterogeneous networking and computation, such as large-scale IoT collaborations or mobile edge computing over trustless IoT environments. Moreover, blockchain provides transparency for MEC networks, where blockchain allows the copy of data records to replicate across edge nodes for public validation, which in return enhances data integrity. This feature is particularly suitable for MEC ecosystems where openness and fairness are required. For example, blockchains can offer transparent ledger solutions to support open and secure data delivery and payment for EDs in a fashion such that EDs can trace and monitor transactions.}

	\textcolor{black}{Moreover, each ED joins the block mining process to maintain the operation of blockchain in MEC. The key purpose of mining is to verify the data transactions, aiming to guarantee the security for the involved edge networks. Accordingly, in the blockchain-based MEC system, EDs perform the mining, and data blocks are secured and chained via an immutable ledger. With more devices mining the blockchain, the security of the edge network increases accordingly.}
	
	
	\subsection{Related Works}
	Recently, many edge task offloading solutions have been proposed \cite{5,6,7}, but these works mostly considered offloading scenarios with a single agent using traditional convex optimization tools. Deep reinforcement learning (DRL) techniques such as deep Q-learning (DQN) have emerged as a promising alternative, by modelling the offloading problem as a Markov decision process (MDP) with using deep neural network (DNN) for function approximation \cite{nguyendeep2021,10,9}. However, these works only used a single agent to handle the entire offloading process which could not work well in large-scale distributed MEC environments. An interesting alternative is to use multi-agent-DRL (MA-DRL) \cite{TMC1} for supporting intelligent task offloading in MEC networks \cite{11}. The work in \cite{13} proposed a non-cooperative MA-DRL scheme where EDs could build their offloading policy independently. Another study in \cite{14} also suggested an MA-DRL approach for joint data offloading and resource allocation in multiple independent edge clouds. Furthermore, a multi-agent Q-learning algorithm was developed in \cite{15} for a joint computation offloading and resource allocation scheme in edge computing. 
	
	\textcolor{black}{ In terms of reputation-based DPoS mining design, the work in \cite{consensus2}  focused primarily on addressing the secure block verification issues in the DPoS mechanism using contract theory. The paper in \cite{add5} suggested a fair voting scheme for the DPoS mechanism via vague set theory. Specifically, this work leveraged a general model of transforming vague sets into fuzzy sets to calculate the comprehensive evaluation indices for agent node selection. The paper in \cite{add6} proposed a contract theory-based optimization scheme for transaction relaying and DPoS based block verification. The authors of this work focused on formulating two mathematical models: value of transaction relaying and value of block verification, and developed the optimal contract to maximize utility of the miners. Further, the work in \cite{add7} proposed a lightweight blockchain-based information trading framework to model the interactions between traffic administration and vehicles in the reputation-based DPoS mechanism via a budgeted auction approach. This study paid attention to the optimization of the mining profit by using a truthful budgeted selection and pricing algorithm. However, a design for low-latency block verification in the reputation-based DPoS mechanism has not been developed. }
	
	Moreover, the research related to task offloading and blockchain mining in MEC networks has been conducted recently.  A blockchain-empowered computation offloading scheme was presented in \cite{11} where blockchain was mainly used for data integrity in the offloading. The authors in \cite{17,18,19} studied edge offloading schemes for blockchain mining tasks with edge clouds, aiming to enhance the quality of service (QoS) for efficient block mining.   Blockchain was also utilized  in \cite{20} to support resource trading for the edge task offloading, while the work in \cite{21} considered a cooperative blockchain-MEC system with an actor-critic DRL algorithm. The study in \cite{related1} optimized the edge computation offloading and resource allocation via a double-dueling deep Q network. The authors in \cite{related2} considered a cooperative computation offloading framework for blockchain-based IoT networks. An MA-DRL algorithm was designed which could allow IoT devices to collaboratively explore the offloading environments in order to minimize long-term offloading costs.  \textcolor{black}{Another work \cite{add8} considered a blockchain-based energy trading scheme to manage the energy trading process toward building a secure energy trading system in Industry 4.0. In \cite{add9}, the problem of resource trading for blockchain-based IoT was studied by using a two-level Stackelberg game with a credit-based payment with smart contracts. Game theory was also applied in \cite{add10} to minimize the economic cost of industrial IoT devices.} However, in most existing works \cite{17,18,19,20,21, related1}, \cite{add8,add9,add10}, the design and optimization of task offloading and blockchain mining were implemented separately, which would result in a sub-optimal performance.
	\begin{table*}
		\scriptsize
		\centering
		\caption{{The comparison of the existing works and our scheme.}}
		\begin{tabular}{|p{3.8cm}||c|c|c|c|c|c|c|c|c|}
			\hline
			\centering \multirow{2}{*}{\textbf{Design features}} &
			\multicolumn{9}{c|}{\textbf{Schemes}} \\
			&	\cite{10}&	\cite{11}&	\cite{13}&	\cite{14}&	\cite{21}&\cite{related1}& \cite{related2}&\cite{add10}	&Our scheme \\
			\hline
			Task offloading with blockchain  &	&	\checkmark&	&	&	\checkmark&	\checkmark&\checkmark &\checkmark	&	\checkmark \\ 
			\hline
			Intelligent edge task offloading 	&	\checkmark&	&	\checkmark&	\checkmark&	\checkmark&	\checkmark& \checkmark&	\checkmark&	\checkmark
			\\  \hline
			Cooperative  edge task offloading 	&	&	&	&	&	&&\checkmark	&\checkmark	& \checkmark
			\\  \hline 
			Lightweight blockchain design&		&	&	&	&	&	&&	&	\checkmark
			\\  \hline
			Joint offloading and mining design &	& &	&	&	&	&&	&	\checkmark \\ 
			\hline
		\end{tabular}
		\label{table:FeatureComparisons}
		\vspace{-0.1in}
	\end{table*}
	\subsection{Motivations and Our Key Contributions}	
	Despite the recent research efforts in blockchain-MEC designs, there are several limitations in existing works, as highlighted below: 
	\begin{itemize}
		\item 	\textcolor{black}{In distributed blockchain-based MEC systems, traditional single-agent DRL algorithms like DQN  \cite{10}, \cite{9}, \cite{17}, \cite{21}, \cite{8} face critical challenges caused by diversified and time-varying local environments. In more details, during the training process of DQN, each agent only observes its local information and cannot know the updates from other agents due to non-collaboration. This
			makes it hard to ensure the stability and convergence of the agents' algorithm \cite{16}. Moreover, this breaks the Markov properties required by the Q-learning algorithm and thus, DQN may not be capable of learning the cooperative offloading policies of EDs. Moreover, the non-cooperative multi-agent DRL solutions \cite{13,14,15} may not be able to learn the cooperative policy; and thus resource usage over the edge network is not efficient which limits the overall offloading performance, e.g., offloading utility. }
		
		\item	{In addition, in most existing blockchain-based MEC schemes \cite{17,18,19,20,21}, the design and optimization of task offloading and blockchain mining are done separately, which would result in sub-optimal performance.} Moreover, the problem of high network latency caused by blockchain mining in the edge offloading system has not been addressed so far \cite{related1, related2}, \cite{add9}, \cite{add10}. To improve the overall performance, a joint offloading and blockchain design is needed for realizing efficient blockchain-based MEC systems. 
	\end{itemize}
	
	\textcolor{black}{Motivated by the aforementioned limitations, we propose a novel cooperative task offloading and blockchain mining (TOBM) scheme for blockchain-based MEC systems enabled by a new MA-DRL solution. Different from existing works \cite{18,19,20,21, related1}, \cite{add8,add9,add10}, we here focus on maximizing the overall system utility as the sum of offloading utility and  mining utility. More specifically, each ED handles data tasks collected from its IoT sensors and deals with block mining tasks simultaneously. To reduce the network latency caused by the blockchain integration in the MEC system, we design a new Proof-of-Reputation (PoR) mining mechanism enabled by a lightweight block verification solution. In particular, we develop a novel distributed DRL-based algorithm using a multi-agent deep deterministic policy gradient (MA-DDPG) approach to optimize the overall system utility.} \textcolor{black}{The proposed MA-DDPG approach enables the efficient learning of the mutual policy among cooperative EDs in the dynamic environment and high-dimensional system state space. Indeed, the proposed MA-DDPG scheme allows EDs to learn mutually the cooperative offloading and mining policy which helps enhance the computation efficiency and thus improves the system utility.} {To enhance the convergence performance in model training and solve the nonstationary issues caused by the concurrently learning process of all EDs in the multi-agent environment, a centralized learning and decentralized execution solution is adopted. As such, the proposed MA-DRL algorithm is first trained at the centralized MEC server, and the learned model is then executed at EDs in a distributed manner.} 
	\textcolor{black}{In fact, the benefits of the MA-DDPG algorithm in edge computing have been proved in recent works for MEC-based industry 4.0 \cite{22}, smart ocean federated learning IoTs \cite{23}, and smart grid \cite{24}. However,  its potential in blockchain-MEC system has  not  been explored so far.} The comparison of our paper and the related works via some key features is summarized in Table~\ref{table:FeatureComparisons}.  In a nutshell, the unique contributions of this article are highlighted as follows:
	
	\begin{enumerate}
		\item 	{We propose a novel cooperative TOBM scheme in a blockchain-based MEC system to enable a joint design of task offloading and blockchain mining for improving the overall system utility.} 
		\item The details of task offloading are presented, where EDs cooperatively offload their IoT data tasks to the MEC server. Moreover, we propose a new PoR mining mechanism enabled by a lightweight block verification strategy, in order to solve latency issues caused by the blockchain adoption in the MEC system. 
		\item 	In the TOBM scheme, each ED as an intelligent agent to learn cooperatively policies, by jointly considering offloading decision, channel selection, transmit power allocation, and computational resource allocation with respect to both offloading and mining states for maximizing the system utility. Then, we propose a novel distributed DRL-based approach using an MA-DDPG algorithm to solve the proposed problem {based on a centralized learning and decentralized execution strategy.}
		\item	\textcolor{black}{ We further develop a game-theoretic solution to model the competition among EDs in offloading and mining as a potential game. We then analyze the properties of the formulated game and prove the existence of a pure Nash equilibrium (NE).}
		\item 	We conduct extensive  numerical simulations and compare with the existing schemes to verify the effectiveness of the proposed scheme. 
	\end{enumerate}
	\subsection{Paper Organization}
	The remainder of this paper is organized as follows. Section~\ref{Blockchain_Offloading} introduces the system model along with the analysis of network model edge task offloading model. The PoR blockchain consensus mechanism is proposed in Section~\ref{consensus}. Based on the offloading and mining design, a joint system utility problem is formulated in Section~\ref{SystemUtility} which is then modelled by a cooperative  offloading game. A new MA-DRL algorithm is proposed to solve the formulated offloading game by using an MA-DDPG algorithm. The simulation results are provided in Section~\ref{simulation} and the comparison with other related offloading schemes is also discussed. Finally, Section~\ref{Conclusion} concludes this article and highlights possible future directions. 
	
	\section{System Model}
	\label{Blockchain_Offloading} 
	In this section, we introduce the network model of the blockchain-based MEC system, and then present the task offloading model. 
	\subsection{Network Model}
	We consider a cooperative TOBM architecture in the blockchain-based MEC system  as illustrated in Fig.~\ref{Overview}. The BS is equipped with an MEC server to provide computation services for EDs.  We denote the set of EDs as $\mathcal{N} = \{1,2,..., N\}$. For the sake of simplicity, we assume that each ED $n \in \mathcal{N}$ has an IoT data task $Y_n$ to be executed \cite{5,6}, which can be defined by a tuple $Y_n = (C_n, D_n, \tau_n), n \in \mathcal{N}$. \textcolor{black}{Herein, $C_n$ denotes the total computational resource (i.e., the number of the CPU cycles)} to accomplish the task $Y_n$. Also, $D_n$ expresses the size of the input data, and $\tau_n$ specifies the maximum permissible latency to accomplish task $Y_n$. In addition to the task offloading function, each ED also participates in the block mining by using a PoR consensus mechanism. The key network components of the blockchain-based MEC system are described as follows:
	\begin{figure}
		\centering
		\includegraphics[width=0.99\linewidth]{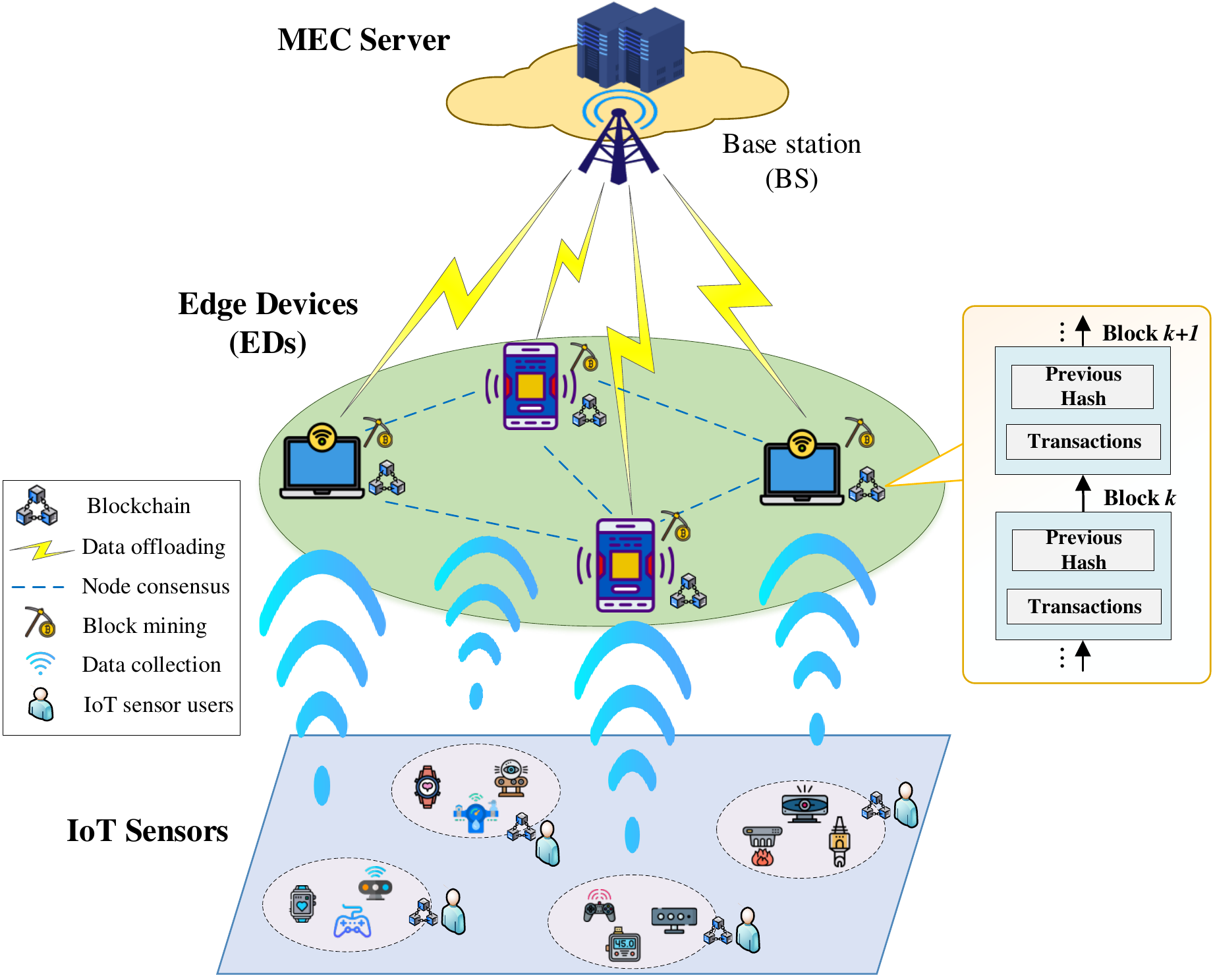} 
		\caption{The cooperative  task offloading and block mining architecture in the blockchain-based MEC system. }
		\label{Overview}
		\vspace{-0.1in}
	\end{figure}
	
	\begin{itemize}
		\item \textit{IoT Sensors:} IoT sensors such as cameras, smart meters, and wearables are responsible for sensing physcial environments, e.g., entertainment, logistics, transportation and healthcare monitoring, and generating data which need to be computed to serve end users. IoT sensors can act as lightweight blockchain nodes to securely communicate and transmit data to their nearby ED via the blockchain network. 
		\item \textit{Edge Devices:} Each ED such as a laptop or a powerful smartphone manages a group of IoT sensors under its coverage. Based on the QoE requirements, EDs can use their computational capability to process data tasks locally or offload to a nearby MEC server via wireless links. EDs also participate in block mining, i.e., transaction verification and block generation, via a PoR consensus mechanism where IoT sensor users vote to select representative EDs to run the mining process. The details of our blockchain mining design are explained in Section~\ref{consensus}. 
		\item \textit{MEC Server:} In our considered blockchain-based MEC system, there is a single MEC sever located at a BS to handle computationally extensive data tasks offloaded from EDs. By analyzing the task profile such as  task sizes, channel conditions, and available resource, EDs can make offloading decisions so that the MEC server can allocate its resources for computation under QoE requirements. 
		\item \textit{Blockchain:} A blockchain network is deployed over the MEC system where each ED acts as a blockchain miner. \textcolor{black}{In this paper, we propose a PoR framework and focus on analyzing the block verification latency that is a significant factor in evaluating the efficiency of a blockchain network. }  The use of our proposed PoR scheme allows EDs to participate in the block mining with an enhanced mining utility which contributes to the system utility improvement in the blockchain-based MEC system. 
		
	\end{itemize}
	
	\begin{figure}
		\centering
		\includegraphics[width=0.8\linewidth]{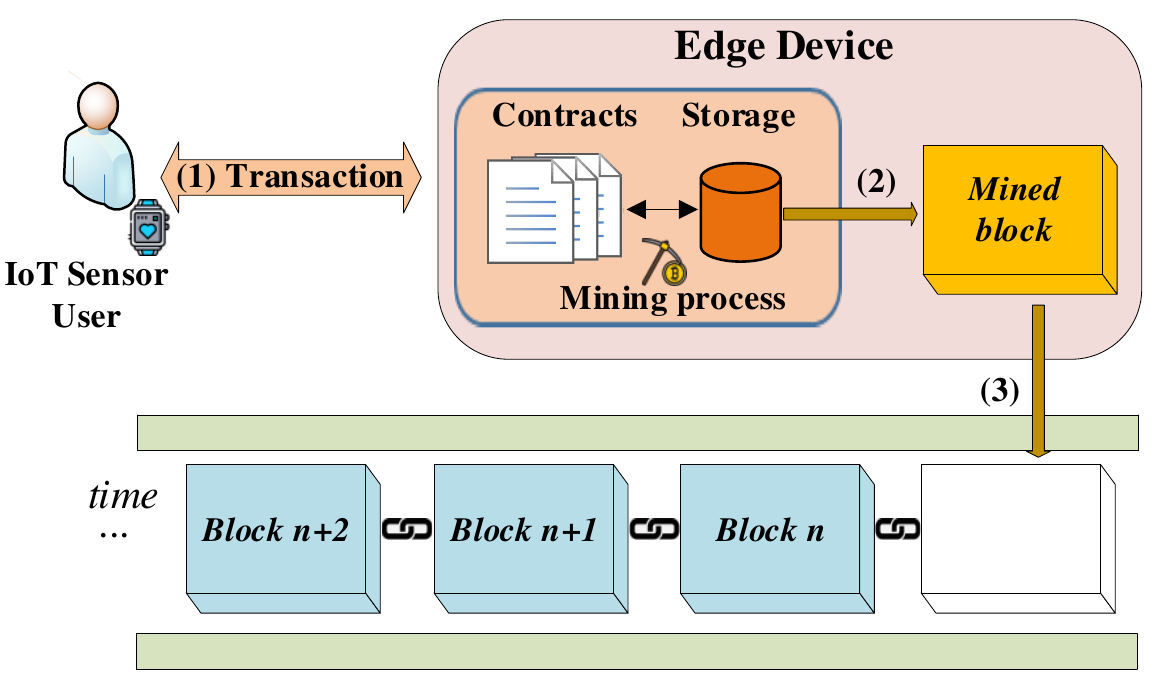} 
		\caption{{\color{black}Illustration of the blockchain operation in our MEC network: (1) An IoT sensor user sends a transaction to its associated ED for triggering the IoT data task offloading, (2) A new block is created to represent the verified transaction, (3) A mined block is appended to the blockchain. }}
		\label{Blockchain_operation}
		\vspace{-0.1in}
	\end{figure}
	{\color{black}The blockchain framework for our MEC network is illustrated in Fig.~\ref{Blockchain_operation}, and its operational concept is explained via the following  steps: 
		\begin{itemize}
			\item \textit{Step 1:} The IoT sensor user first creates a transaction with metadata (i.e. user ID), user signature and timestamp from its wallet account and sends it to the associated ED. The user then submits a transaction to the ED for a certain request, such as IoT data task offloading.  
			\item \textit{Step 2:} The ED releases its available resources and processes the request from the user. For example, an ED can use a smart contract \cite{nguyensecure2021}, a self-executing software running on blockchain, to automatically perform transaction authentication, user verification, or resource trading. Moreover, the ED collaborates with other mining members to aggregate the transactions offloaded from IoT users to build a block after a certain period of time. Then, the EDs participate in the mining to verify the block using a consensus mechanism, e.g., PoR.
			\item \textit{Step 3:} After the mining, if all miners achieve an agreement on the verified block, this block with its signature is then appended to the chain of blocks in  chronological order. Finally, all network entities receive this block and synchronize the copy of the blockchain.
	\end{itemize}}
	
	\subsection{Edge Task Offloading Model}
	\label{Offloadingmodel}
	Here, we present the communication model and the computing model for the edge task offloading. 
	
	\subsubsection{Communication Model}
	We denote $\mathcal{K}  = \{1,..., K\}$ as the set of available sub-bands at the BS. We define a task offloading policy, which also incorporates the uplink sub-band scheduling, by a binary variable $x_{n}^k$,  ($n \in \mathcal{N}, k \in \mathcal{K}$). Here, $ x_{n}^k = 1$ indicates that the task $Y_n$ from ED $n$ is offloaded to the MEC server via sub-band $k$, and $ x_{n}^k = 0$ otherwise. Each computation task can be either executed locally at the ED or offloaded to the MEC server under a feasible offloading policy: 
	\begin{equation}  
	\sum_{k \in \mathcal{K}} x_{n}^k \leq 1, n \in \mathcal{N}.
	\vspace{-0.06in}
	\end{equation}
	In line of the above discussion, we define the task offloading policy $\boldsymbol{X}$ that contains all the task offloading variables $x_{n}^k$ as $\boldsymbol{X} = \{x_{n}^k| x_{n}^k = 1, n \in \mathcal{N}, k \in \mathcal{K}\}$. 
	Besides, we denote $\mathcal{N}_n = \{n \in \mathcal{N}| \sum_{k \in \mathcal{K}}x_{n}^k=1\}$ as the set of EDs offloading their tasks to the MEC server. 
	
	Moreover, we consider that each ED and the BS have a single antenna for uplink communications. We denote $h_{n}^k$ as the uplink channel gain between the ED $n$ and the BS on sub-band $k$. Let $\boldsymbol{P} = \{ p_n^k | 0 < p_n^k \leq P_n^k, n \in \mathcal{N}_n \}$ denote the transmit power policy of EDs, where $ p_n^k $ is the transmit power of ED $n$ when offloading the task $Y_n$ to the BS via the channel $k$, subject to a maximum budget $ P_n^k$. \textcolor{black}{We also assume that the MEC system has a total operational frequency band $B_{MEC}$ that is divided into $K$ sub-bands of an equal size $W = B_{MEC}/K$ [Hz].} Then, the transmission data rate of the ED $n$ can be calculated as
	\begin{equation} 
	R_{n} = Wlog_2 \left(1+\frac{p_n^k h_{n}^k}{\sigma^2 + \sum_{j \in  \mathcal{N}_n, j \neq n}\left(x_j^k p_j^k h_j^k\right)}\right),
	\end{equation}
	where  $\sigma^2$ is the background noise variance and the second term at the denominator is the interference among mobile users in the same channel. We also denote $ x_{n}= \sum_{k \in \mathcal{K}}x_{n}^k, \forall n \in \mathcal{N}$. Thus, the required time that the ED $n$ upload its task input $D_n$ via the uplink is specified as
	\begin{equation} 
	T_n^{up} =  \frac{D_n}{R_{n}}, \forall n \in \mathcal{N}. 
	\end{equation}
	Accordingly, the energy consumption of ED $n$ for offloading the task $Y_n$ is specified as  
	\begin{equation} 
	E_n^{up} = p_nT_n^{up} = p_n \frac{D_n}{R_{n}}, \forall n \in \mathcal{N},
	\end{equation}
	where $p_n = \sum_{k \in \mathcal{K}} p_n^k, \forall n \in \mathcal{N}$. 
	
	\subsubsection{Computing Model}
	We consider two computing modes: local execution and edge offloading.
	
	- \textbf{Local execution:}
	Let $f_n^l$ denote the computational resource of ED $n$ (in CPU cycles/s) allocated to execute the data task, which should not exceed its total computation capacity $F_n$. Thus, we can define the policy of computational resource allocation of EDs as  $\boldsymbol{F} = \{ f_n^l | 0 < f_n^l \leq F_n, n \in \mathcal{N} \}$. The time consumed to execute the task input $D_n$ \textcolor{black}{(with $C_n$  in CPU cycles)} at an ED $n$ is expressed as
	\textcolor{black}{\begin{equation}
		T_n^l = \frac{C_n}{f_n^l}.
		\end{equation}} 
	Moreover, the energy consumption of an ED $n$ when executing its task locally is specified as 
	\begin{equation}
	E_n^l = \kappa (f_n^l)^2 C_n,
	\end{equation}
	where $\kappa$ is the energy coefficient depending on the chip architecture \cite{5} and $C_n$ is the CPU workload of ED $n$.
	
	- \textbf{Edge offloading:}
	For the offloading case, the MEC server at the BS can provide computation services to multiple EDs concurrently. Compared to the local device, the MEC server has much more powerful computation capacity $f^e$ (in CPU cycles/s) and more stable power supply. The execution time of task $Y_n$ at the MEC server can be calculated as
	\begin{equation} 
	T_n^{ex} = \frac{C_n}{f^e}, \forall n \in \mathcal{N}.
	\end{equation}
	Similar to \cite{5,6}, we do not model the downloading part due to the small size of data results compared to the offloading data. 
	
	In summary, the latency cost consumed by the ED $n$ when offloading its task $Y_n$ is given by
	\begin{equation} 
	T_n^{off} = T_n^{up} + T_n^{ex} = \left(\frac{D_n}{R_{n}} + \frac{C_n}{f^e} \right), \forall n \in \mathcal{N}. 
	\end{equation}
	
	Moreover, the energy cost consumed by the ED $n$ when offloading its task $Y_n$ is only associated with the data transmission, which is given by
	\begin{equation} 
	E_n^{off} = E_n^{up} =  p_n \frac{D_n}{R_{n}}, \forall n \in \mathcal{N}.
	\end{equation}
	\section{Blockchain Consensus Design}
	\label{consensus}
	In the blockchain-based MEC system, a crucial component is blockchain consensus that aims to mine the blocks of transactions (i.e., IoT data records) and add them to the blockchain. To handle the transactions, the EDs work as blockchain miners to perform mining. In the blockchain-based edge offloading environment, latency is one of the most important factors determining the efficiency of a blockchain system. Given a consensus algorithm, when the number of transactions to the blockchain increases, the consensus workload to validate and append them into the blockchain will increase significantly. In current consensus schemes, e.g., DPoS \cite{sun2020joint}, each miner node must implement a repeated verification process across the miner network, which results in unnecessary consensus latency and network bandwidth waste. A possible solution is to reduce the number of miner nodes to reduce the consensus latency, but it potentially compromises the security of blockchain because of the high probability of adding compromised transactions from malicious nodes \cite{consensus1}. To solve these mining issues, here we propose a new lightweight Proof of Reputation (PoR) consensus mechanism for our blockchain system. Compared to the DPoS scheme, we make an improvement in the miner selection based on a reputation score evaluation approach. Moreover, instead of using a repeated verification among miner nodes, we implement a lightweight block verification solution that allows each miner only needs to verify once with another node during the consensus process, which would significantly reduce the verification latency and save network bandwidth. There are two main parts of our PoR consensus, including miner node selection and block verification, as illustrated in Fig.~\ref{Fig:PoR}.
	\subsection{Miner Node Selection}
	In this phase, the IoT users first calculate the reputation score of EDs and then select the miner nodes to implement the mining process.
	
	\subsubsection{Reputation Calculation} 
	\label{Section:MinerNodeSelection}
	In our MEC system, IoT sensors' users participate in the delegate selection process to vote the mining candidates among EDs for performing blockchain consensus. In this regard, each IoT user votes its preferred ED with the most reputation. Here, the reputation of an ED is measured by its mining utility with respect to mining latency. That is, an ED exhibits a lower mining latency will have a better mining utility which increases its reputation. To this end, we define a mining utility function of each ED as:
	\begin{equation}
	\label{MiningUtility}
	J_n^{mine} = \left[e^{1-\frac{T_n^{PoR}}{\tau_n}} -1\right]^+,
	\end{equation}
	where $T_n^{PoR}$ is the mining latency of the ED $n$ (its detail will be explained in the following sub-section), $\tau_n$ denotes the task execution latency constraint. $[y]^+ = \max \{y,0\}$ implies that the reputation of an ED is set to 0 if the mining latency $T_n^{PoR}$ is exceeded to its task execution constraint $\tau_n$.  
	
	\subsubsection{Miner Selection}
	Based on the calculated reputation score, each sensor user will vote for the ED candidates as the miners based on their reputation ranking. The top EDs with highest reputation scores are selected to become edge miners (EMs) to perform mining, as indicated in Fig.~\ref{Fig:PoR}. \textcolor{black}{Also, similar to the traditional DPoS framework \cite{sun2020joint}, in our PoR mechanism, each of the active EMs takes turn to act as a block manager during its time slot to coordinate the consensus process. In other words, there is one manager in each consensus process. In the next time slot, another active EM will undertake this manager role. }
	\subsection{Lightweight Block Verification}
	\begin{figure}
		\centering
		\includegraphics [width=0.9\linewidth]{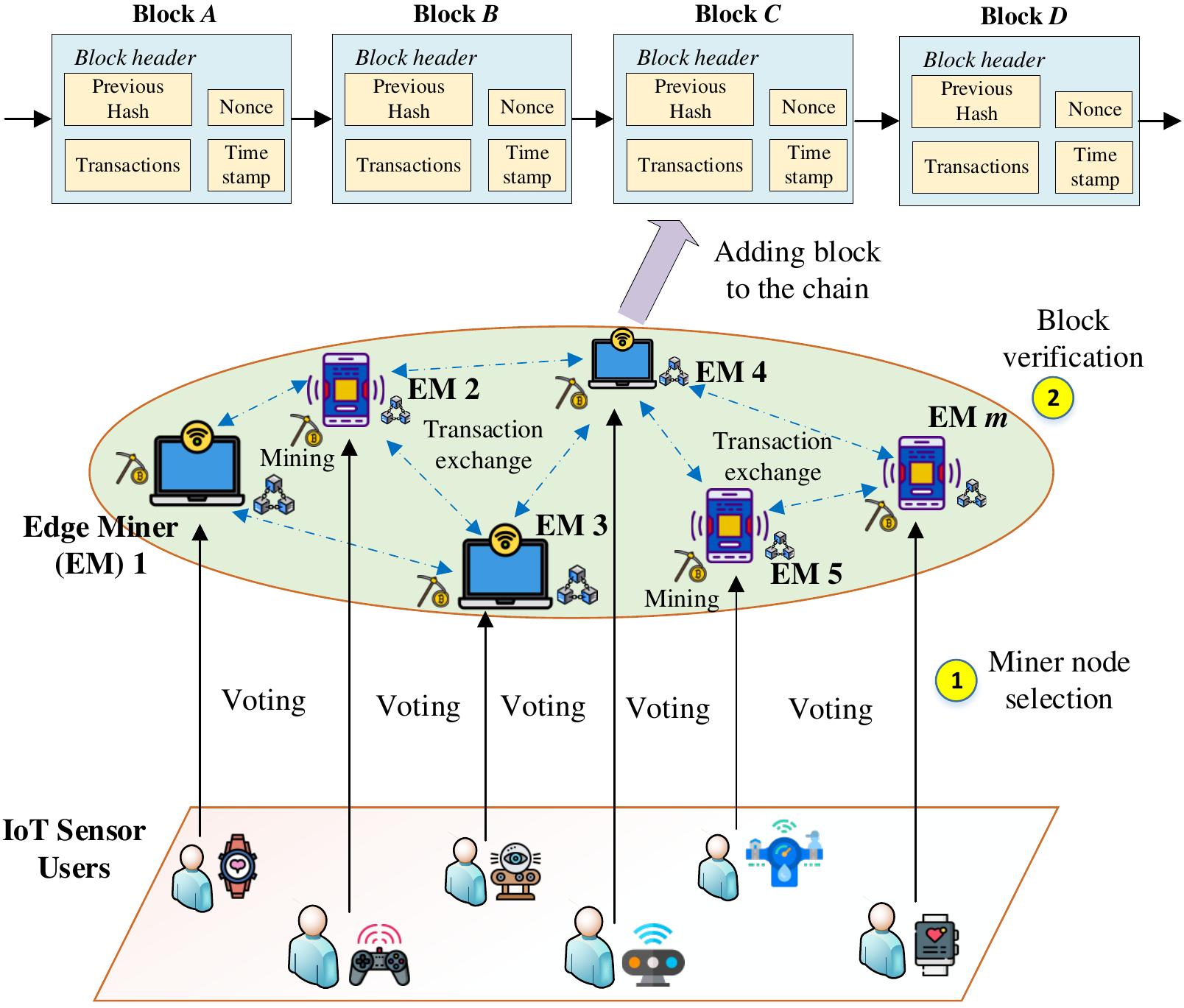}
		\caption{The proposed PoR consensus in our blockchain-based MEC system. }
		\label{Fig:PoR}
		\vspace{-0.1in}
	\end{figure}
	In this phase, the block manager first produces an unverified block $B$ that contains several offloading transactions collected in a given amount of time. Then, the manager broadcasts this created block to all EMs within the miner network for verification. Different from the traditional DPoS scheme \cite{sun2020joint} which relies on a repeated verification process among miners, here we implement a lightweight verification solution that allows each miner to only verify once with another node during the consensus process. \textcolor{black}{Algorithm~\ref{Al:PoR1} presents how our proposed block verification procedure is performed. In lines \ref{blockdevide} to \ref{blockdevide1}, the block manager divides the block $B$ consisting of transactions into $N$ transaction parts $Tr_n$, $n \in \mathcal{N}$ that will be assigned to each mining member $EM$ within the miner group. Each miner $EM$ will also be  assigned a unique random number $R_n$. In  lines \ref{verify} to \ref{verify1}, an $EM$ selects any miner $s$ ($s \in \mathcal{N} \backslash  n$) within the miner group to implement the verification for its assigned transaction part $Tr_n$. If 51\% of the EMs respond positive verification, and the sum of random numbers $Sum$ calculated by all EMs is equal to the initial number set $Rnd$, the block manager accepts the verified block $B'$ and adds it to the blockchain with a signature. For instance, in Fig.~\ref{Fig:PoR}, the EM $4$ works as a manager to create the block $C$ and append it into the blockchain. Otherwise, the manager discards it from the network (in lines \ref{append1} to  \ref{append2}).}
	
	\begin{algorithm}
		\footnotesize
		\caption{\textcolor{black}{Procedure of the proposed PoR consensus}} 
		\label{Al:PoR1}
		\begin{algorithmic}[1]
			\STATE \textbf{Input:}  the unverified block $B$, a set of EMs $\mathcal{N}$
			\STATE \textbf{Output:} the verified block $B'$
			\STATE \textbf{Initialization:} Select an unverified block $B$, group the selected EMs $EM$ in the list $Array[n]$, $n \in \mathcal{N}$, initiate public key $Array[n].PK$ and block manager $BM$
			\STATE Divide the block $B$ into $N$ parts $Tr_n$, $Sum \leftarrow 0$ \label{blockdevide}
			\FOR {$n = 1,..., N$}
			\STATE Set a part of block $Tr_n \rightarrow Array[n].content$
			\STATE Assign a random number $R_n \leftarrow Random()$
			\STATE Calculate a signature as $Sig_n \leftarrow Hash(Array[n].content,Array[n].PK, timestamp)$
			\ENDFOR  \label{blockdevide1}
			\STATE Specify the total random number $Rnd = N(1+\frac{N-1}{2})$
			\FOR {$n = 1,..., N$} \label{verify}
			\STATE Run a random function $s = Random.randrange(1,N,1)$
			\IF {$s \neq n$}  
			\STATE \textit{Select a random different EM within the list}
			\STATE Send the $Tr_n$ to $EM_s$: $EM \rightarrow EM_s: (Tr_n, Array[n].PK, Sig_n, timestamp)$
			\STATE Verify the transaction $Tr_n$
			\IF {($Array[n].PK_{EM_s} == EM_s^{EM.PK}) \cap (Verify(Sig_n) \leftarrow true$)}
			\STATE $Sum \leftarrow Sum + R_n$
			\ENDIF
			\ENDIF 
			\ENDFOR  \label{verify1}
			\IF {$Sum == Rnd$} \label{append1}
			\STATE Accept the block $B$ as a verified one ($B'$)  and send it back to the block manager $BM$
			\STATE The manager $BM$ appends the verified block $B'$ into the blockchain network: $BM \rightarrow *: (BM_{PK},B', Sign_B, timestamp)$
			\ELSE
			\STATE Discard the block $B$ from the blockchain
			\ENDIF \label{append2}
		\end{algorithmic}
	\end{algorithm}
	
	\subsection{Latency of Block Verification} In this sub-section, we calculate the verification latency incurred by the mining. For simplicity, we assume that the transaction part $Tr_n$ (which also expresses the size) is the same for all EMs. Each EM is willing to allocate a certain CPU resource $\phi_n$ (in CPU cycles) for the verification of transaction part $n$. Then, the CPU resource allocation policy for block verification of EDs can be defined as  $\boldsymbol{G} = \{ \phi_n | 0 < \phi_n \leq \Phi_n, n \in \mathcal{N} \}$, where $\Phi_n$ is the CPU resource budget of the ED $n$. Further, the size of verified transaction result for the $Tr_n$ is denoted by $Tr_n^{re}$. Hence, the transaction verification task can be expressed as a tuple ($Tr_n, \phi_n, Tr_n^{re}$). 
	
	Conceptually, the block verification process in our proposed PoR mechanism at an EM experiences four steps: (1) unverified block transmission from the block manager to the EMs, (2) local block verification at the EM, (3) broadcasting of the verification result among two EMs,  and (4) transmission of verification result feedback from the EMs to the manager. For a miner EM $n$, the time required to complete these steps is expressed as:
	\begin{equation}
	\label{EquationPoR}
	T_n^{PoR} = \frac{Tr_n}{r_n^d} + \frac{\phi_n}{F_n} +\xi Tr_n|L^2|  + \frac{Tr_n^{re}}{r_n^u}, n \in \mathcal{N},
	\end{equation}
	where $r_n^u$ and $r_n^d$ are uplink and downlink transmission rates between the miner $n$ and the block manager. Here, the transmission time of an unverified transaction part $Tr_n$ from the block manager to the miner is $ \frac{Tr_n}{r_n^d}$, while the local verification time of this transaction is $ \frac{\phi_n}{F_n}$. Moreover, similar to \cite{consensus2}, the time for transaction broadcasting among two miners is a function of transaction size $Tr_n$ and network scale $Tr_n|L^2|$ (which means two miners for transaction verification), which is defined as $\xi Tr_n|M^2|$. Here, $\xi$ is a pre-defined parameter of broadcasting verification result and comparison among two miners, which can be acquired from the previous verification records \cite{consensus2}. Besides, $ \frac{Tr_n^{re}}{r_n^u} $ is the verification feedback time 
	
	Meanwhile, in the traditional DPoS scheme \cite{sun2020joint}, each miner has to implement a repeated verification process among all miners for the block $B$, instead of dividing into separate transaction parts like our proposed PoR model. Therefore, the verification latency of the DPoS consensus at an  EM $n$ is expressed as \cite{consensus2}:
	\begin{equation}
	\label{EquationDPoS}
	T_n^{DPoS} = \frac{B}{r_n^d} + \frac{\phi_n^B}{c^B_n} +\xi B|L^N|  + \frac{B^{re}}{r_n^u},
	\end{equation}
	where $\phi_n^B$ is the CPU resource occupied to verify the block $B$ under the computation budget $c_n^B$. $B^{re}$ denotes the size of verified result of the block $B$. $|L^N|$ expresses the whole miner network which means all miners $n$ join the repeated block verification in each consensus process, instead of two-miner verification in our PoR scheme. 
	By comparison of equations~\ref{EquationPoR}~and~\ref{EquationDPoS}, it can be seen that the proposed PoR scheme needs less time for block verification in comparison with the traditional DPoS scheme, for the same block size and number of miners. Moreover, our mining scheme can save much network bandwidth due to less message exchange during the consensus process. The benefits of our proposed PoR mechanism are verified in the following sections. 
	
	\section{System Utility Formulation and Proposed MA-DRL Algorithm}
	\label{SystemUtility}
	In this section, we present the system utility formulation for our proposed TOBM scheme based on the joint consideration of offloading utility and mining utility as presented in the previous sections. Then, we derive the system utility optimization problem as a cooperative  game and propose a new MA-DRL algorithm to solve it.  
	\subsection{System Utility Formulation}
	In this paper, we formulate the system utility for the TOBM scheme by taking both offloading utility and mining utility into account. 
	\subsubsection{Offloading Utility}
	Here, we focus on formulating the QoE-aware offloading utility function. In an MEC system, the offloading's QoE is mainly characterized by their task computation time, i.e., $T_n$ and energy consumption, i.e., $E_n$. Specifically, $T_n$ and $E_n$ can be specified as
	\begin{equation} 
	T_n = T_n^{off}x_{n} + T_n^l(1-x_{n}), \forall n \in \mathcal{N},
	\end{equation}
	\begin{equation} 
	E_n = E_n^{off}x_{n} + E_n^l(1-x_{n}), \forall n \in \mathcal{N}.
	\end{equation}
	Accordingly, we define a QoE-aware utility function to measure the offloading utility that is specified as a trade-off between the time and energy consumption of the task compared with local execution
	\begin{equation} 
	J_n^{off} = \lambda^T_n \left(\frac{T_n^l-T_n}{T_n^l}\right) + \lambda^E_n \left(\frac{E_n^l-E_n}{E_n^l}\right), 
	\end{equation}
	where $\lambda^T_n,\lambda^E_n \in [0,1]$ (with $\lambda^T_n+\lambda^E_n =1, \forall n \in \mathcal{N}$) are set by ED $n$ to show the preference on time and energy cost when computing the task $Y_n$. If the task is emergency, the ED can increase the weighting factor of time consumption. Meanwhile, if the ED is operating with low battery, the factor of energy consumption should be preferred. 
	
	It is noting that here, the offloading utility function $J_n^{off}$ reflects the improvement in QoE over local execution, that is measured by $ \left(\frac{T_n^l-T_n}{T_n^l}\right)$ and $ \left(\frac{E_n^l-E_n}{E_n^l}\right)$, respectively. Specifically, when the ED $n$ executes the task locally, the offloading utility equals 0 (i.e., $ J_n^{off} = 0$). If the computation cost (i.e., latency and energy consumption) of the offloading mode is lower than that of the local execution mode, the offloading utility ($J_n^{off}$) can be positive, which indicates the offloading's QoE improvement. However, if offloading too many tasks, the EDs can suffer from higher latency due to the traffic congestion, which would reduce the QoE. As a result, the offloading utility ($J_n^{off}$) can be negative. 
	
	To this end, we formulate the QoE-aware offloading utility for the MEC system. Given the offloading decision policy $\boldsymbol{X}$, the transmission power policy $\boldsymbol{P}$, and the computational resource allocation policy  $\boldsymbol{F}$, we define the offloading utility as the weighted sum of all MDs' offloading utilities $J_n^{off}$, denoted as: 
	\begin{equation}
	J^{off} = \sum_{n \in \mathcal{N}} J_n^{off}(\boldsymbol{X}, \boldsymbol{P},\boldsymbol{F}),
	\end{equation}
	\subsubsection{Mining Utility}
	We adopt the mining utility built in equation \ref {MiningUtility} to analyse the efficiency of the mining in the blockchain-enabled task offloading. Given the CPU resource allocation policy $\boldsymbol{G}$, each MEC server yields an utility $J_n^{mine}$ when performing the mining process. Then, we can specify the total mining utility of EDs in the MEC system as 
	\begin{equation}
	J^{mine} = \sum_{n \in \mathcal{N}} J_n^{mine}(\boldsymbol{G}).
	\end{equation}
	
	Accordingly, the system utility of each MD $J_n$ can be expressed as 
	\begin{equation}
	\label{equa:utility-idividual}
	J_n = J^{off}_n + J^{mine}_n. 
	\end{equation}
	
	\subsubsection{System Utility Formulation}
	{In this paper, our objective is maximize the total system utility as the sum of the offloading utility and the mining utility for the proposed TOBM scheme:}
	\begin{subequations}
		\label{Equa:Optimization}
		\begin{align} 
		& \underset{\boldsymbol{X},\boldsymbol{P},\boldsymbol{F},\boldsymbol{G}}{\maximize} 
		&& J^{off} + J^{mine} \label{Objective}\\
		& \phantom{P1} \text{subject to} 
		&& x_{n}^k \in \{0,1\}, \forall n \in \mathcal{N}, k \in \mathcal{K}, \label{constraint1} \\
		&&& \sum_{k \in \mathcal{K}} x_{n}^k \leq 1, n \in \mathcal{N}, \label{constraint2}\\
		&&& 0 < p_n^k \leq P_n^k, \forall n \in \mathcal{N}_n, k \in \mathcal{K}, \label{constraint3}\\
		&&& 0 < f_n^l \leq F_n, \forall n \in \mathcal{N}, \label{constraint4}\\
		&&& 0 < \phi_n \leq \Phi_n, \forall n \in \mathcal{N}_n \label{constraint5}\\
		&&& T_n \leq \tau_n, \forall n \in \mathcal{N}. \label{constraint6}
		\end{align}
	\end{subequations}
	
	Here, the constraints \eqref{constraint1} and \eqref{constraint2} imply that each task can be either executed locally or offloaded to the MEC server via a sub-channel. \eqref{constraint3} shows the transmission power constraint of each ED. The constraint \eqref{constraint4} states that each ED $n$ must allocate a positive computational resource to execute the computing task, but not exceed the total computation budget $F_n$. Each ED $n$ also must allocate a positive CPU resource for the block verification under a maximum CPU capability $\Phi_n$, as indicated in \eqref{constraint5}. Constraint \eqref{constraint6} ensures that each data task needs to be completed under a delay threshold. 
	
	\textcolor{black}{The key intuition behind this integrated calculation is that in the blockchain-based MEC system, each ED needs to simultaneously perform task offloading and block mining. In this context, the evaluation of the system quality, e.g., overall utility performance, must consider both offloading utility and mining utility. Indeed, an ED needs to minimize its offloading latency and energy consumption to maintain the quality of task offloading service, while also minimizing its mining latency to maintain the quality of block mining service. Therefore, we come up with a final solution to satisfy both services, aiming to optimize the overall system utility performance. }
	
	The problem~\ref{Equa:Optimization} is non-convex and centralized. As a dynamic TOBM problem (due to varying channel conditions and task sizes) and high-dimensional system state space (due to the increase of EDs), the use of traditional optimization  approaches results in high computational complexity which would hinder the applicability of the proposed model in practical blockchain-based MEC scenarios. Moreover, most of current solutions \cite{8,9,10} rely on single-agent learning which suffers from some critical shortcomings:
	\begin{itemize}
		\item 	\textit{Dimensionality:} The cardinalities of DNN input and output are generally proportional to the number of EDs, and thus the use of centralized learning to obtain the optimal policy for all EDs is challenging. Moreover, exploration in high-dimensional state space is inefficient especially when the number of EDs increases exponentially, which makes the learning in the blockchain-based MEC system impractical. 
		\item 	\textit{Information transmission:} Since the centralized learning always requires full information of all EDs (e.g., data task state, resource state, block size state) for decision making, the information transmission becomes challenging with the increase of EDs. 
	\end{itemize}
	Due to the powerlessness of centralized learning algorithms in the multi-agent environment like our considered blockchain-based MEC system, we propose to use a distributed MA-DRL scheme to solve our cooperative TOBM problem, as presented in the following sub-sections. 
	\subsection{Cooperative  Learning Formulation}
	First, we convert the objective function \ref{Objective} from a system utility maximization problem to a reward maximization problem. To do this, we formulate the task offloading  problem using a multi-agent version of  MDP, also known as a \textit{Markov game} which is denoted by a tuple $<\mathcal{N, \mathcal{S}, \mathcal{A}, \mathcal{O}}>$. Here, each ED $n$ is considered as an intelligent agent to learn its optimal policy by observing the TOBM environment and collaborating with other agents, aiming to achieve the optimal system utility. Then, we have $\mathcal{N} = \{1,2,..., N\}$ as the set of EDs (or agents). Moreover, $\mathcal{S} = \{s_1,s_2,..., s_N\}$ is defined as the set of states,  $\mathcal{A} = \{a_1,a_2,..., a_N\}$ is a set of agent actions, and $O = \{o_1,o_2,..., o_N \}$ denotes a set of observations for agents. We assume that the considered cooperative TOBM scheme operates on discrete time horizon with each time slot $t$ equal and non-overlapping, and the communication parameters keep unchanged during each time slot. Now we define each item in the tuple at each time slot $t$ as follows.
	\subsubsection{State} The environment states in the cooperative TOBM network include five components: task state $S_{task}(t)$, channel state $S_{channel}(t)$, power state $S_{power}(t)$, resource state $S_{res}(t)$, and transaction state $S_{trans}(t)$. Therefore, the system state is defined as a matrix:
	\begin{equation} 
	\mathcal{S}(t) = \{S_{task}(t), S_{channel}(t), S_{power}(t),S_{res}(t),S_{trans}(t)\},
	\end{equation}
	where each state vector is explained as follows. $S_{task}(t)$ is defined as $S_{task}(t) = [D_n(t), C_n(t)], n \in \mathcal{N}$ where $D_n(t)$ represents the computation task size of the ED $n$ and $C_n(t)$ is the required input CPU cycles number to complete the task data $D_n(t)$. $S_{channel}(t)$ is defined as: 
	\begin{equation} 
	S_{channel}(t) = c^k_n(t) = 
	\begin{bmatrix}
	& c_{1,1} & \cdots & c_{1,K} \\
	& \vdots  & \ddots & \vdots  \\
	& c_{N,1} & \cdots & c_{N,K} 
	\end{bmatrix},
	\end{equation}
	where $c^k_n(t)$ indicates whether the sub-channel $k$ is used by ED $n$ at time slot $t$. If yes, $c^k_n(t) =1$, otherwise $c^k_n(t) =0$. Also, $S_{power}(t)$ is defined as:
	\begin{equation} 
	S_{power}(t) = p^k_n(t) = 
	\begin{bmatrix} \footnotesize
	& p_{1,1} & \cdots & p_{1,K} \\
	& \vdots  & \ddots & \vdots  \\
	& p_{N,1} & \cdots & p_{N,K} 
	\end{bmatrix},
	\end{equation}
	where $p^k_n(t)$ represents the ED $n$'s transmit power level in the $k$th sub-channel, which is a continuous variable and satisfies $0 < p^k_n(t) \leq P^k_n$. Moreover, the resource state $S_{res}(t)$ is expressed as 
	\begin{equation}
	S_{res}(t) = \{v_1(t), v_2(t),..., v_N(t)\},
	\end{equation}
	where $v_n(t)$ contains the states of current available computational resource $f^l_n(t)$ and CPU resource $\phi_n(t)$ of the ED $n$. Lastly,  the transaction state $S_{trans}(t)$ is defined as 
	\begin{equation}
	S_{trans}(t) = \{Tr_1(t), Tr_2(t),..., Tr_N(t)\},
	\end{equation}
	where $Tr_n(t)$ is the transaction state of ED $n$. 
	\subsubsection{Action} By observing the system states, each ED needs to make actions in each time step to deal with the task execution and block mining, including offloading decision, channel selection, transmit power selection, computational resource allocation, and CPU resource allocation. Accordingly, the action space can be expressed as 
	\begin{equation}
	\mathcal{A}(t) = \{x_{n}^k(t),k(t),p^k_n(t), f^l_n(t), \phi_n(t)\},
	\end{equation}
	where each action compopent is explained as follows:
	\begin{itemize}
		\item Offloading decision $x_{n}^k(t)$: $x_{n}^k(t) \in \{0,1\}$, ($n \in \mathcal{N}, k \in \mathcal{K}$). Each ED $n$ makes decision to execute the task locally $x_{n}^k(t)=0$ or offload it to the MEC server $x_{n}^k(t)=1$ via the channel $k$, based on the current task state $S_{task}(t)$.
		\item Channel selection $k(t)$: $k(t)=[1,2,...,K]$. Each ED $n$ selects one of the available channels to offload the task to the MEC server, based on the current channel state $S_{channel}(t)$.
		\item Transmit power selection $p^k_n(t)$: $p^k_n(t) \in (0,P_n^k]$, ($ n \in \mathcal{N}_n, k \in \mathcal{K}$). Each ED $n$ chooses a transmit power value to transmit the data task to the MEC server with respect to the current task state $S_{task}(t)$ and channel state $S_{channel}(t)$. 
		\item  Computational resource allocation $f^l_n(t)$: $f^l_n(t) = [f^l_1(t),f^l_2(t),..., f^l_N(t)]$. Each ED $n$ allocates part of its computational resource to execute the task with respect to the current resource state $S_{res}(t)$ and task state $S_{task}(t)$. 
		\item CPU resource allocation $\phi_n(t)$: $\phi_n(t) =[\phi_1(t),\phi_2(t),...,\phi_N(t)]$. Each ED $n$ allocates part of its CPU resource to verify the blockchain transaction, based on the current resource state $S_{res}(t)$ and transaction state $S_{trans}(t)$. 
	\end{itemize}

	\subsubsection{System Reward Function} The system reward at one time slot $t$ is the sum of the rewards of all EDs. Each ED $n$ will get a reward $r(s_n(t),a_n(t))$ in a certain state $s_n(t)$ after executing each possible action $a_n(t)$. In our paper, the system reward function should be positively correlated to the objective function in the optimization problem \ref{Equa:Optimization}, aiming to maximize the system utility of all EDs. Then, we can specify the system reward function of our offloading network at each time slot $t$ as
	\begin{equation}
	\label{Equation_reward} 
	r(s(t),a(t)) = \sum_{n \in \mathcal{N}}r(s_n(t),a_n(t)) = J(t),
	\end{equation}
	where $J(t)= J^{off}(t) + J^{mine}(t)$ is the total system utility of the blockchain-based MEC system. 
	\begin{figure}
		\centering
		\includegraphics[width=0.99\linewidth]{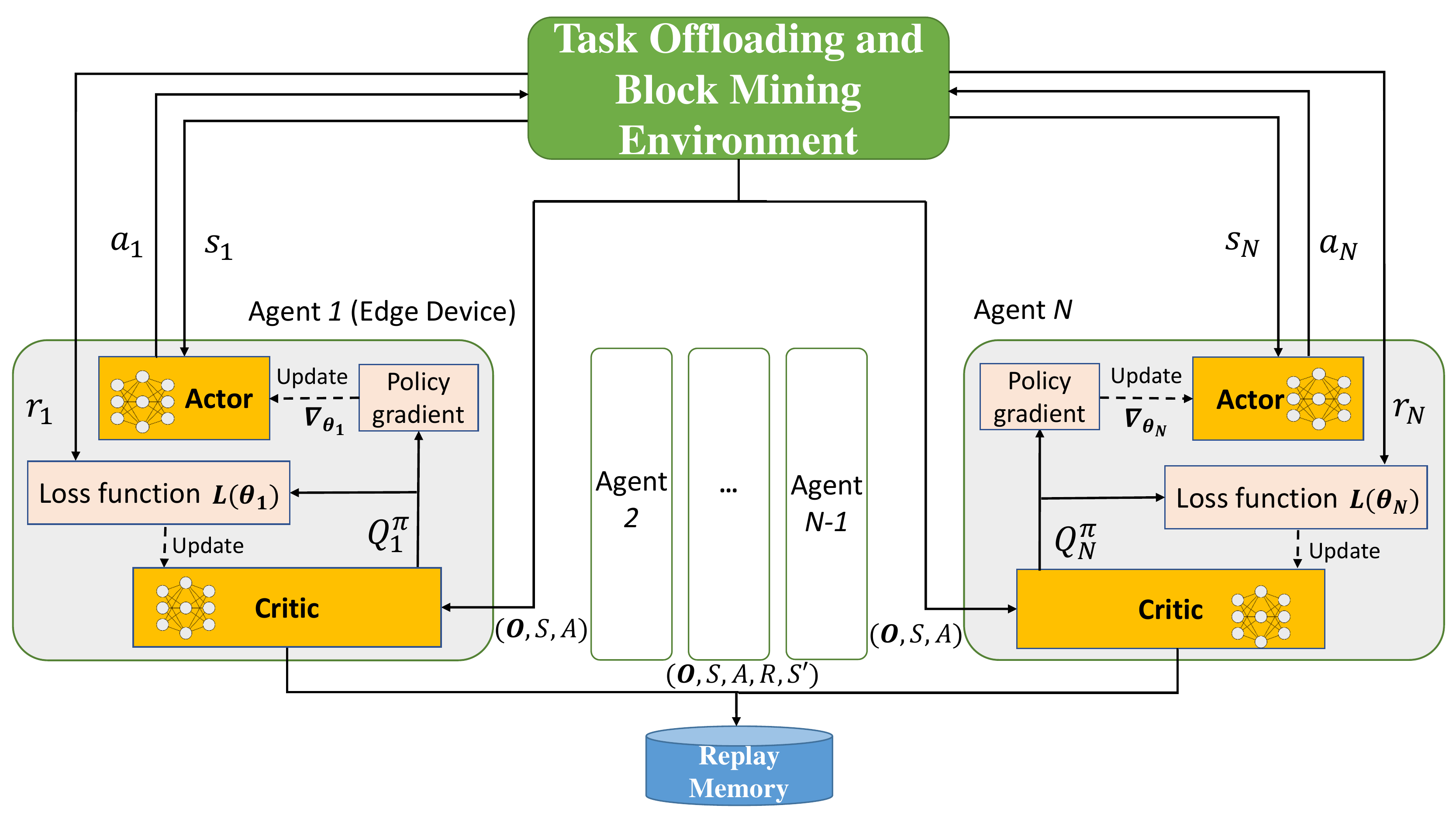} 
		\caption{ The proposed MA-DDPG architecture. }
		\label{MA-DRL}
		\vspace{-0.1in}
	\end{figure}
	\subsection{Proposed MA-DRL Algorithm for Cooperative TOBM}	
	
	In the cooperative TOBM problem in our blockchain-based MEC system, conventional single-agent \cite{8,9,10} or independent multi-agent \cite{13,14,15} solutions  are unable to obtain the cooperative policies of EDs due to the nonstationary and partially observable environment. Indeed, when policies of other agents change (i.e., due to computation mode preference), the ED (agent) $n$ observation $O_n$ can be changed (nonstationary), which makes the obtained reward $r_n(t)$ different from the accumulated reward from its actual state-action pair. Moreover, in independent multi-agent learning schemes, the agent $n$ only has the local information and cannot know the updates from other agents due to non-collaboration. This would affect the agent $n$'s reward $r_n(t)$ and make the learning algorithms hard to ensure stable convergence \cite{lowe2017multi}.
	{Therefore, we adopt a centralized learning and decentralized execution solution to implement our MA-DRL algorithm for the proposed TOBM scheme. }
	\subsubsection{Preliminaries of Reinforcement Learning}
	In RL, an agent takes some actions to obtain rewards through the trial and error procedure according to a predefined MDP, aiming to accumulate experience as much as possible to construct an optimal policy. Specifically, the state-action function can be updated using the agent $n$'s experience tuple $(s_n(t),a_n(t), r_n(t), s_n(t+1))$ at each time step $t$ as 
	\begin{equation}
	Q(s_n(t),a_n(t)) \leftarrow Q(s_n(t),a_n(t)) + \alpha \sigma(t),
	\end{equation}
	which is called as the Q-learning algorithm \cite{8}, where $\sigma(t)$ is a temporal difference (TD) error that would be zero for the optimal Q-value, $\alpha$ is the learning rate, and $\gamma$ is the discount factor between $(0, 1)$. 
	\subsubsection{Proposed MA-DDPG Algorithm}
	Here, we present an MA-DRL approach using an MA-DDPG algorithm to solve the cooperative TOBM problem in the blockchain-based MEC system, as illustrated in Fig.~\ref{MA-DRL}. Different from RL, DRL uses a DNN as the non-linear approximator to sample the loss function at each training step in order to alleviate the computational complexity for the large-scale offloading problem. 
	Here, agents cooperatively offload their tasks to the MEC server and perform mining to form a shared learning environment consisting of all EDs and the MEC server. {In the centralized training step, the information of state-action of all EDs is aggregated by the MEC server to train the DRL model where each agent can obtain the global view of the learning environment to learn collaboratively with other agents. This makes the learning environment stationary and thus enhances the convergence performance. After training at the MEC server, the learned parameters are downloaded to each of EDs to execute the model for decision making based on its own locally observed information.}
	
	We denote $\pi = \{\pi_1, \pi_2,...,\pi_N\}$ as the set of all agent policies and $\theta = \{\theta_1, \theta_2,...,\theta_N\}$ as the parameter set of corresponding policies. Every agent updates its parameters $\theta_n$ to obtain the optimal policy $\pi^*_{\theta_n} = argmax_{\theta_n}J(\theta_n)$, where $J(\theta_n)$ is the objective function (also the expected reward) of agent $n$ as defined in equation~\ref{Equation_reward}. 
	{MA-DDPG is a deterministic policy gradient-based off-policy \textit{actor-critic} operating over continuous action spaces in a multi-agent environment.} {Here, the \textit{actor} generates deterministic action $a$ over time slots with a behavior network and the \textit{critic} evaluates the behavior of the \textit{actor} with a target network.} In the training, the \textit{actor} updates the behavior network by computing the gradient of the objective function $J(\theta_n)$ as
	\begin{equation}  
	\bigtriangledown_{\theta_n} J(\pi_n)= \mathds{E}_{\textbf{o},a\sim\mathcal{D}}\left[\bigtriangledown_{\theta_n}Q_n^\pi(\textbf{o}, a_1,...,a_N). \bigtriangledown_{\theta_n}\pi_n(a_n|o_n) \right],
	\end{equation}
	with ($\textbf{o} = \{o_1,..., o_N \}$) as the observation set, $Q_n^\pi(\textbf{o}, a_1,...,a_N)$ is a centralized action-value function of the agent $n$ with $a_1,a_2,...,a_N$ as the actions of all agents and is learned separately for each $n \in \mathcal{N}$. Also, $\mathcal{D}$ is the memory buffer for experience replay, containing multiple episode samples ($\textbf{o},a,r,\textbf{o}'$). 
	Moreover, the \textit{critic} updates the behavior Q-function $Q_n^\pi()$ in a fashion that minimizes the loss function, which is written as 
	\begin{equation}  
	L(\theta_n) = \mathds{E}_{\textbf{o},a,r,\textbf{o}'}\left[( y_n-  Q_n^\pi(\textbf{o}, a_1,a_2,...,a_N) )   ^2\right],
	\vspace{-0.06in}
	\end{equation}
	where $y_n = r_n+\gamma Q^\pi_n(\textbf{o}',a_1',a_2',...a_N')|_{a'_n=\pi'_n(o_n)}$ is the is TD target and $\pi'_n(o_n)$ defines the target policies with delayed parameters $\theta'_n$. {The training procedure is summarized in Algorithm~\ref{Al:DDPG}.} \textcolor{black}{Here, the procedure consists of two main phases, the planning phase and the updating phase. In the planning phase, we use an $\epsilon$-greedy policy to balance the exploration and exploitation for updating the Q function (line~\ref{Algo:start}). At each time epoch, each ED executes an action and estimates the system reward, i.e., system utility, and stores training information in the replay memory (lines~\ref{Algo:training}-\ref{Algo:store}). After each action, the ED moves to the next step, updates the critic and actor networks as well as corresponding target networks (lines~\ref{Algo:update1}-\ref{Algo:update2}). The training is iterated until achieving the desired system reward performance. }
	
	We can see that the update of $\theta_n$ of the target network in the policy gradient method $\bigtriangledown_{\theta_n}$ would guide how the agent ED acts correctly to obtain the optimal policy. It is based on the fact that that the value of DNN in the target actor network is fixed for several iterations, and the weights $\theta_n$ of the DNN in the actor behavior network are updated. In other words, all agent EDs in the blockchain-based MEC system can maximize their expected function $J(\theta_n)$ (i.e., user utility) and obtain stable policies even via interactions between EDs and the environment. This makes the learning environment stationary even when the policies $\pi_n$ change, which would enhance the quality of policy evaluation for improving the overall system utility. After the training is completed, EDs download the learned policy network parameters from the MEC server and update the target network parameters for the actor and critic as
	\begin{equation}
	\label{Equa:ActorUpdate}
	\theta_j' \leftarrow \zeta\theta_j + (1-\zeta) \theta_j',
	\end{equation}
	where $\zeta \in (0,1)$ is the update step. 
	\subsubsection{{Computational  Complexity Analysis}}
	In our MA-DRL algorithm, the training process is implemented in the MEC server with sufficient computational resource. Therefore, we mainly focus on the computational  complexity of the execution process at EDs. Here, a DRL agent with a DNN is established for each ED and all DNNs run in parallel across the MEC network. As a result, the overall complexity of the multi-agent system can be determined by the complexity of a single DNN at an ED. We assume that are $K$ neurons at the input layer of the DNN for each ED, and $Z$ as the number of neurons at the output layer. Also, the hidden layer is $L$, and the number of neurons at hidden layers is $H$. Accordingly, the computation cost at a DNN is $(KH+(L-1)HH+HZ)$ =$ \mathcal{O}(H(K+(L-1)H+Z))$. Also, the complexity of using activation function is $ \mathcal{O}(HL)$. Hence, the total complexity is $ \mathcal{O}(H(K+HL-H+Z+L))$ which can be simplified as $ \mathcal{O}(H(K+HL+Z))$.
	\begin{algorithm}
		
		\caption{\textcolor{black}{The MA-DDPG training procedure in the blockchain-based MEC system}}
		\begin{algorithmic}[1]
			\label{Al:DDPG}
			\STATE \textbf{Input:}  Replay memory $\mathcal{D}$, time budget $T$, exploration probability $\epsilon$, discount factor $\gamma$, update step $\zeta$
			\STATE \textbf{Output:} The optimal policy $\pi^*_{\theta_n}$ and maximum reward $r^*(s,a)$ 
			\STATE \textbf{Initialization:} Initialize the deep Q network  $Q(s,a)$ with random weight $\theta$ and $\theta'$, initialize the exploration probability $\epsilon \in (0,1)$
			\FOR{episode = 1,..., \textit{M}} 
			\STATE Initialize the state $s_0 \leftarrow\{S_{task}(t), S_{channel}(t), S_{power}(t)\}|_{t=0}$
			\FOR{$t = 1,2,...,T$}
			\STATE For each agent ED $j \in \mathcal{N}$, select a random action $a_j(t)$ with probability $\epsilon$, otherwise $a_j(t) = \pi_{\theta_j}(s_j(t))$ \label{Algo:start}
			\STATE Execute actions $a(t) = (a_1(t), a_2(t),...,a_N(t))$ by performing offloading decision $x_{n}^k(t)$, channel selection $k(t)$, transmit power selection $p^k_n(t)$, computational resource allocation $f^l_n(t)$, and CPU resource allocation $\phi_n(t)$ \label{Algo:training}
			\STATE Observe the system reward $r(t)$ via~\ref{Equation_reward} and the new state $\textbf{s}'$
			\STATE Store $(s(t), a(t), r(t), \textbf{s}'(t))$ into the memory $\mathcal{D}$ \label{Algo:store}
			
			\FOR{$agent ~ j=1 $ to $ N $} \label{Algo:update1}
			\STATE Sample random mini-batch of transitions ($s_j,a_j,r_j,\textbf{s}'_j$) from  $\mathcal{D}$
			\STATE Set $y_j = r_j+\gamma Q^\pi_j(\textbf{s}'_j,a_1',a_2',...a_N')|_{a'_j=\pi'_j(o_j)}$
			\STATE Update behavior critic by minimizing the loss: $L(\theta_j) = \frac{1}{S}\sum_{j}\left[y_j-  Q_j^\pi({s}_j, a_1,a_2,...,a_N)    \right]^2$
			\STATE Update actor by using the sampled policy gradient:
			$\bigtriangledown_{\theta_j} J(\pi_j)= \frac{1}{S}\left[\bigtriangledown_{\theta_j}Q_j^\pi(\textbf{s}_j', a_1,a_2,...,a_N). \bigtriangledown_{\theta_j}\pi_j(a_j|s_j) \right]$
			\ENDFOR
			\STATE Update the target network parameters for each agent via~\ref{Equa:ActorUpdate} \label{Algo:update2}
			\ENDFOR 
			\ENDFOR 
		\end{algorithmic}
	\end{algorithm}
	\vspace{-0.1in}
	
	{\color{black}\subsection{Cooperative Game-theoretic Solution }
		We next develop a game-theoretic approach for the proposed TOMB problem, where each ED can act as a game player to react to other players' decisions for maximizing its utility \cite{add1}, \cite{add2}. After a number of steps, all the EDs self-organize into a mutual equilibrium state, i.e., the Nash equilibrium, at which no ED can further increase its utility by unilaterally altering its strategy. 
		
		We first define the game as $G = \{ \mathcal{N}, \{\mathcal{A}_n\} _{n \in \mathcal{N}},\{J_n\} _{n \in \mathcal{N}} \}$, where  $\mathcal{N}$ is the set of rational game players, $\mathcal{A}_n$ is the strategy set for player $n$, and $J_n$ is the utility function of ED $n$. Let denote $a_n$ as the offloading decision profile of the player $n$ over the wireless sub-bands $\mathcal{K}$, from \ref{equa:utility-idividual} we can rewrite as
		\begin{equation}
		\label{equa_game1}
		J_n(a_n) =  J_n^{off} I_{\{a_n=1\}} + J_n^{mine},
		\end{equation}
		where  $I_{z}$  is an indicator function. If $z$ is true, $I_{z} = 1$; otherwise,  $I_{z} = 0$. Based on \ref{equa:utility-idividual} and \ref{equa_game1}, the utility of a player $n$ in the cooperative game can be expressed as
		\begin{equation}
		J(a_n, a_{-n})  = \begin{cases}
		J_n^{mine}, & a_n = 0\\
		J_n^{off}  + J_n^{mine}, & a_n = 1
		\end{cases},
		\end{equation}
		where $a_{-n}$ is the offloading decisions of all players exept $n$. Accordingly, by considering the influence of other players on the utility optimization of a player $n$ in the cooperative game, we can express the game-theoretic utility function as
		\begin{multline} 
		\scriptstyle
		J(a_n, a_{-n})  =  \begin{cases} \scriptstyle
		J_n^{mine}, & \scriptstyle a_n = 0\\ \scriptstyle
		\sum_{m\#n} \left(J_n^{off}(a_n)+J_m^{off}(a_m)\right) I_{\{a_m=a_n\}} \\ \scriptstyle + J_n^{mine}(a_n), & \scriptstyle a_n = 1
		\end{cases}
		\end{multline}
		where $m \in \mathcal{N} \backslash  \{n\}$. 
		\\
		\textbf{Theorem 1:} \textit{The collaborative game $G = \{ \mathcal{N}, \{\mathcal{A}_n\} _{n \in \mathcal{N}},\{J_n\} _{n \in \mathcal{N}} \}$ has a pure Nash equilibrium (NE) and guarantees the finite improvement property.}
		\\
		\textit{Proof:} We first prove that the proposed game G is an exact (cardinal) potential game with potential function
		\begin{multline}
		\label{equation:potentilgame}
		\Psi(a_n, a_{-n}) = \frac{1}{2} \sum_{n\in \mathcal{N}} \sum_{m\#n} (J_n^{off}+J_m^{off}) I_{\{a_m=a_n\}} I_{\{a_n =1\}} \\ +  \sum_{n\in \mathcal{N}} J_n^{mine} I_{\{a_n =1\}} + \sum_{n\in \mathcal{N}} J_n^{mine} I_{\{a_n =0\}}, 
		\end{multline}
		such that 
		\begin{multline}
		\Psi(a'_n, a_{-n}) - \Psi(a_n, a_{-n}) = J(a'_n, a_{-n}) - J(a_n, a_{-n}), \\ \forall a_n, a_{-n} \in \mathcal{A}_n.
		\end{multline}
		We now consider three cases as follows:
		\\
		Case 1: $a_n > 0, a_n' >0$
		
		\begin{multline}
		\Psi(a_n', a_{-n}) - \Psi(a_n, a_{-n}) \\= \frac{1}{2}\sum_{m\#n} \left(J_n^{off}(a_n') - J_n^{off}(a_m)\right)I_{\{a_m=a_n'\}}  \\+  \frac{1}{2}\sum_{m\#n} \left(J_n^{off}(a_m) - J_n^{off}(a_n')\right)I_{\{a_n'=a_m\}} + J_n^{mine}(a_n')  \\- \frac{1}{2}\sum_{m\#n} \left(J_n^{off}(a_n) - J_n^{off}(a_m)\right)I_{\{a_m=a_n\}} \\-   \frac{1}{2}\sum_{m\#n} \left(J_n^{off}(a_m) - J_n^{off}(a_n)\right)I_{\{a_n=a_m\}} + J_n^{mine}(a_n) - J_n^{mine}(a_n) \\ = 
		\sum_{m\#n} \left(J_n^{off}(a_n')+J_m^{off}(a_m)\right) I_{\{a_m=a_n'\}}   + J_n^{mine}(a_n') \\- \sum_{m\#n} \left(J_n^{off}(a_n)+J_m^{off}(a_m)\right) I_{\{a_m=a_n\}}  - J_n^{mine}(a_n) \\ = J(a'_n, a_{-n}) - J(a_n, a_{-n}).
		\end{multline}
		\\
		Case 2: $a_n = 0, a_n' >0$
		\begin{multline} 
		\Psi(a_n', a_{-n}) - \Psi(a_n, a_{-n}) \\= \frac{1}{2}\sum_{m\#n} \left(J_n^{off}(a_n') - J_n^{off}(a_m)\right)I_{\{a_m=a_n'\}}  \\+  \frac{1}{2}\sum_{m\#n} \left(J_n^{off}(a_m) - J_n^{off}(a_n')\right)I_{\{a_n'=a_m\}} + J_n^{mine}(a_n')  - J_n^{mine} \\ = \sum_{m\#n} \left(J_n^{off}(a_n')+J_m^{off}(a_m)\right) I_{\{a_m=a_n'\}}   + J_n^{mine}(a_n') - J_n^{mine}
		\\= J(a'_n, a_{-n}) - J(a_n, a_{-n}).
		\end{multline}
		Case 3: $a_n > 0, a_n' = 0$
		
		Similar to Case 2, it is also straightforward to prove that $\Psi(a_n', a_{-n}) - \Psi(a_n, a_{-n}) = J(a'_n, a_{-n}) - J(a_n, a_{-n})$. 
		
		Therefore, the game $G$ is an exact potential game with the potential function given in~\ref{equation:potentilgame}. Finally, according to the potential game theory \cite{add1}, our proposed collaborative game $G$ has an NE and possesses the finite improvement property.}
	
	\section{Simulations and Performance Analysis }
	\label{simulation}
	In this section, we perform extensive simulations to verify the performance of the proposed TOBM scheme. 
	
	\subsection{Simulation Setting}
\begin{figure}
		\centering
		\includegraphics [width=0.9\linewidth]{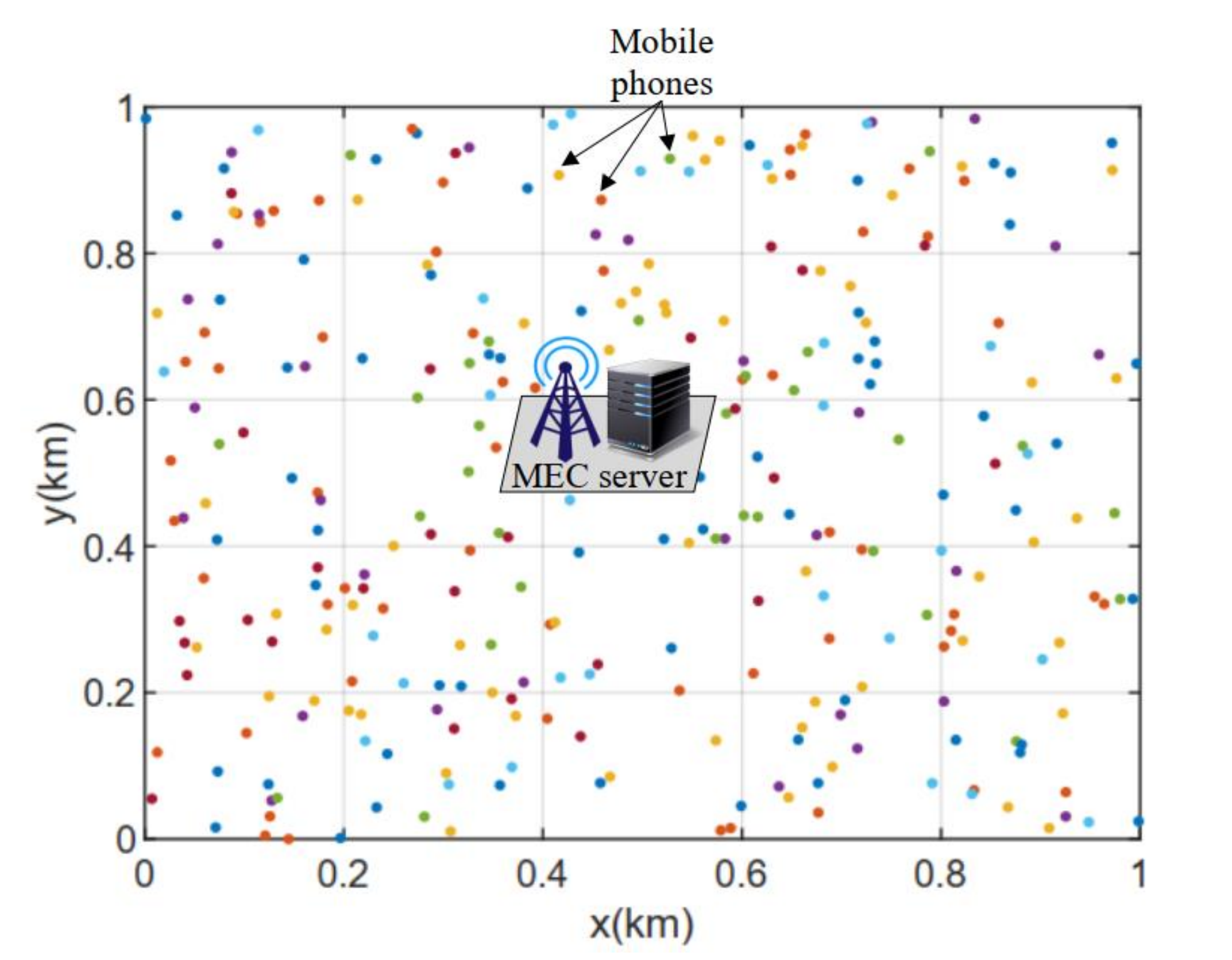}
		\caption{{Illustration of the MEC network with an MEC server and distributed mobile phones.}}
		\label{Fig:Distribution1}
		\vspace{-0.1in}
	\end{figure}

	We here leverage the widely used mobile wireless dataset provided by Shanghai Telecom\footnote{ http://www.sguangwang.com/dataset/telecom.zip} for numerical simulations. We select maximum 500 mobile phones as EDs and an MEC server in a sub-area of Shanghai city with the geographical distribution as illustrated in Fig.~\ref{Fig:Distribution1}.  Moreover, IoT sensor data traces from location services \cite{Dataset1} collected during 6 months in 2014 are selected as data tasks for task offloading simulations. Inspired by \cite{7,8,9,10}, our simulation parameters are configured as in Table~\ref{Table:Simulation}.

\begin{table}
	\centering
	\caption{Simulation parameters.}
	\label{Table:Simulation}
	
	\setlength{\tabcolsep}{5pt}
	\begin{tabular}{c}
		\includegraphics[width=0.99\linewidth]{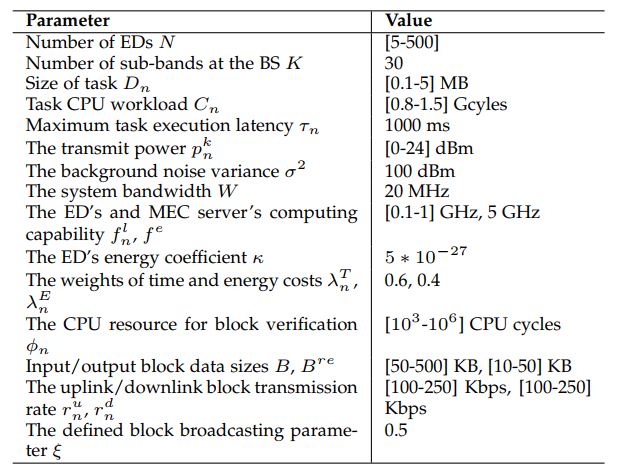} 
	\end{tabular}
	\vspace{-0.16in}
\end{table}

	\begin{figure*}[t!]
		\centering
		\begin{subfigure}[t]{0.5\textwidth}
			\centering
			\includegraphics[width=0.99\linewidth]{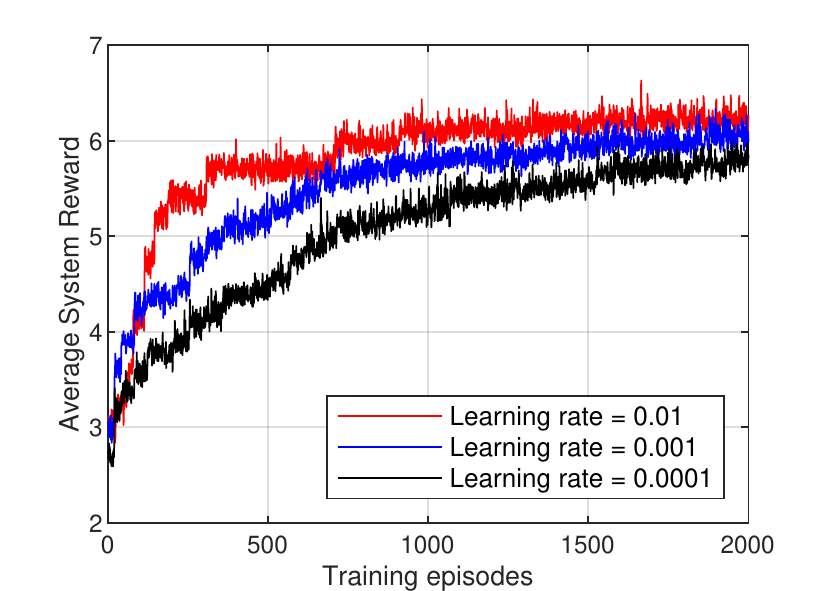} 
			\caption{Comparison of average system rewards with different learning rates. }
		\end{subfigure}%
		~
		\begin{subfigure}[t]{0.5\textwidth}
			\centering
			\includegraphics[width=0.99\linewidth]{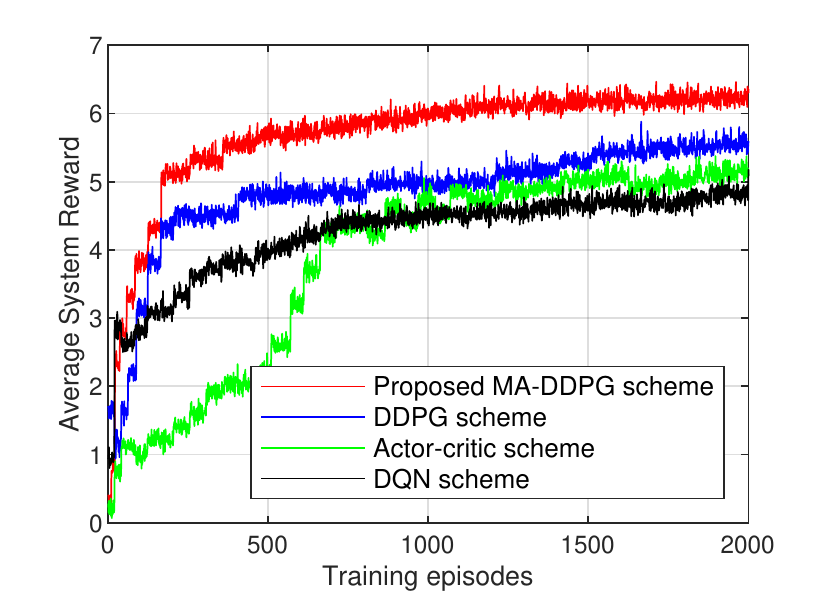} 
			\caption{Comparison of average system rewards with different algorithms.   }
		\end{subfigure}
		\caption{Evaluation of training performance.}
		\label{Convergence_Result}
		\vspace{-0.1in}
	\end{figure*}
	Moreover, the channel gains $h_n^k$ are generated using distance-dependent path-loss model $L[dB] =140.7 + 36.7log_{10}d_{[km]}$. For the proposed multi-agent DRL algorithm, the discount factor $\gamma$ equals 0.85 and the replay memory capacity and training batch size are set to $10^5$ and 128, respectively. The update step $\zeta$ in the critic-actor training is set to 0.8. The used DNN structure has three hidden layers (64, 32 and 32 neurons) with ReLU as the activation function and Adam as the optimizer. For blockchain mining, we set up 10 transactions per block and vary the numbers of mining nodes (i.e., EDs) from 2 to 100. The other mining parameters are included in Table~\ref{Table:Simulation}. The results from simulation are averaged from 50 runs of numerical simulations. 
	
	\subsection{Evaluation of Training Performance}
	We first evaluate the training performance of our proposed algorithm. To prove the advantage of the proposed cooperative MA-DDPG algorithm, we compare its performance with the state-of-the-art non-cooperative schemes, including DDPG, actor-critic \cite{21} and DQN \cite{14}. Here, DDPG and actor-critic are policy-based algorithms where each ED agent only observes the local information and does not the information of other EDs during the training. Meanwhile, DQN is a value-based algorithm where each ED also has no information of other EDs. 
	
	Fig.~\ref{Convergence_Result}(a) shows the performance of average system reward with different learning rates $\alpha$. It can be seen that the learning rate affects the learning rewards over the training episodes. That is, when the learning rate decreases, the convergence performance of the proposed algorithm decreases due to slow learning speed. Based on our experimental results, the learning value $\alpha=0.01$ yields the best reward performance and has good convergence rate and thus we use it in the following system simulations and evaluations.

	Fig.~\ref{Convergence_Result}(b) shows the learning curves of the average system reward with the increase of episodes for an MEC system with 50 EDs. It is clear that our MA-DDPG scheme is more robust and yields the best performance in terms of average system reward, compared to the baseline schemes. This is because the proposed scheme allows EDs to learn mutually the cooperative  policy which helps reduce the channel congestion and user interference, and enhance computational resource efficiency for improving the overall system reward.  Meanwhile, in the DQN and actor-critic schemes, EDs greedily access the wireless channel spectrum to maximize their own utility as much as possible without collaborating with each other, which increases the possibility of channel collision and thus results in higher offloading latency. Consequently, the average system reward becomes worse. Although the DDPG scheme shows a better reward performance than these two schemes, it still remains a non-stationary learning issue and its average reward is lower than that of the MA-DDPG scheme. 
	\subsection{Evaluation of Task Offloading Utility}
	\begin{figure*}[t!]
		\centering
		\begin{subfigure}[t]{0.5\textwidth}
			\centering
			\includegraphics[width=0.99\linewidth]{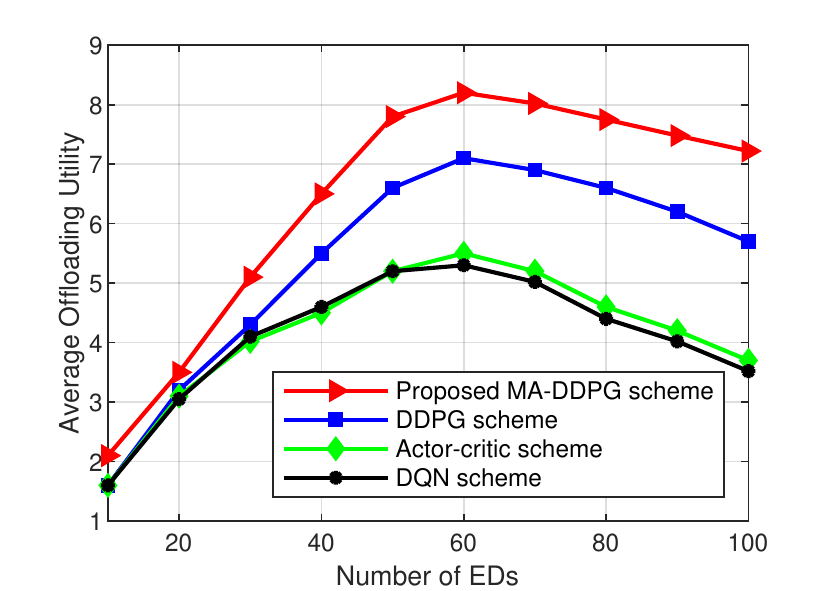} 
			\caption{Average offloading utility with different numbers of EDs. }
		\end{subfigure}%
		~
		\begin{subfigure}[t]{0.5\textwidth}
			\centering
			\includegraphics[width=0.99\linewidth]{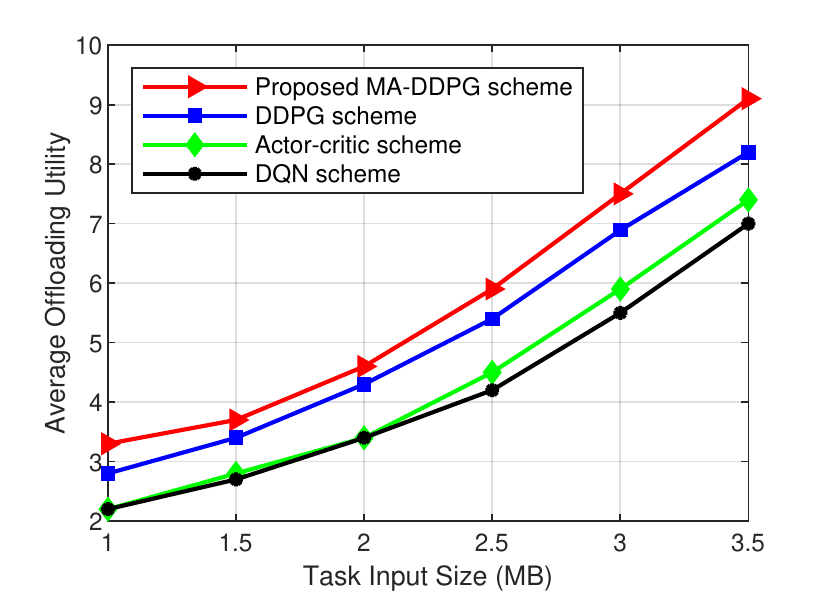} 
			\caption{Average offloading utility with different IoT data task sizes.   }
		\end{subfigure}
		\caption{Evaluation of task offloading performance.}
		\label{Offloading_Performance}
		\vspace{-0.1in}
	\end{figure*}
	
	Next, Fig.~\ref{Offloading_Performance}(a) indicates the performance of average offloading utility versus the different numbers of offloaded EDs. It can be seen that when the number of EDs is small ($<60$), the average offloading utility increases with the number of EDs because in this case, the MEC system can support sufficient spectrum and computing resources to handle all tasks offloaded from EDs. However, after exceeding some thresholds (e.g., $N=60$ EDs), the system utility decreases because the higher the number of offloaded EDs, the higher the competition of resource usage (i.e., channel spectrum). \textcolor{black}{Note that the configured number of sub-bands $K=30$ is relatively low with respect to the increase in the number of EDs, thus the channel bandwidth allocated for each ED decreases when there are more EDs in the system.} In turn, this increases the offloading latency and thus, degrades the overall offloading utility. Nevertheless, our MA-DDPG scheme still achieves the best utility performance due to its cooperative  offloading policies among EDs compared to the other schemes with selfish learning. For instance, in the case of 100 EDs, the average offloading utility of the MA-DDPG scheme is 22.5\%, 37.5\%, and 43.6\% higher than those of the DDPG, actor-critic, and DQN schemes, respectively. These results also imply that as the EDs number increases, the cooperative policy learned by MA-DDPG becomes more important in the cooperative edge task offloading.
	
	Moreover, we evaluate the effect of different task sizes on the average offloading utility as shown in Fig.~\ref{Offloading_Performance}(b). We find that a higher task size results in a higher offloading utility. Specifically, when the task size is relatively high ($>2$MB), the offloading utility increases significantly. The reason is that the edge computation mode becomes more efficient than the local execution mode in terms of lower computing cost in handling the larger-size tasks due to the high MEC capability. This leads to the increase of the user utility $J_n$ which thus enhances the system-inter utility $J$. Particularly, when the task size increases, the advantage of the proposed MA-DDPG scheme over the baselines become more profound, with the larger performance gaps and better utilities. For example, as the task size is 3.5 MB, the proposed MA-DDPG scheme achieves 30\%, 24.5\% and 21.7\% higher offloading utilities compared with the DDPG, actor-critic, and DQN schemes, respectively.
	\subsection{Evaluation of Blockchain Performance}
	Here, we evaluate the performance of our proposed PoR consensus scheme via numerical simulations using Python programming and compare it with the traditional DPoS scheme \cite{sun2020joint} via the verification block latency and bandwidth usage metrics. 
	\begin{figure*}[t!]
		\centering
		\begin{subfigure}[t]{0.5\textwidth}
			\centering
			\includegraphics[width=0.99\linewidth]{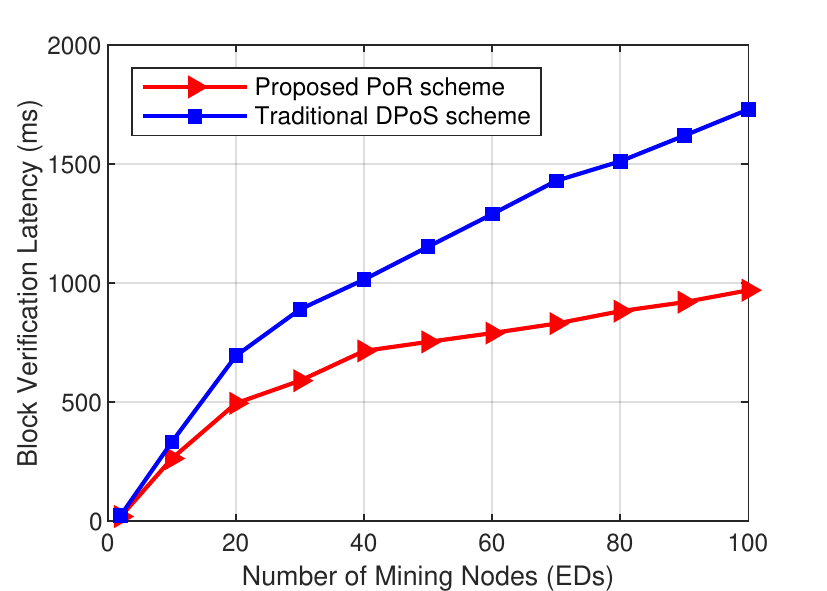} 
			\caption{Comparison of block verification latency. }
		\end{subfigure}%
		~
		\begin{subfigure}[t]{0.5\textwidth}
			\centering
			\includegraphics[width=0.99\linewidth]{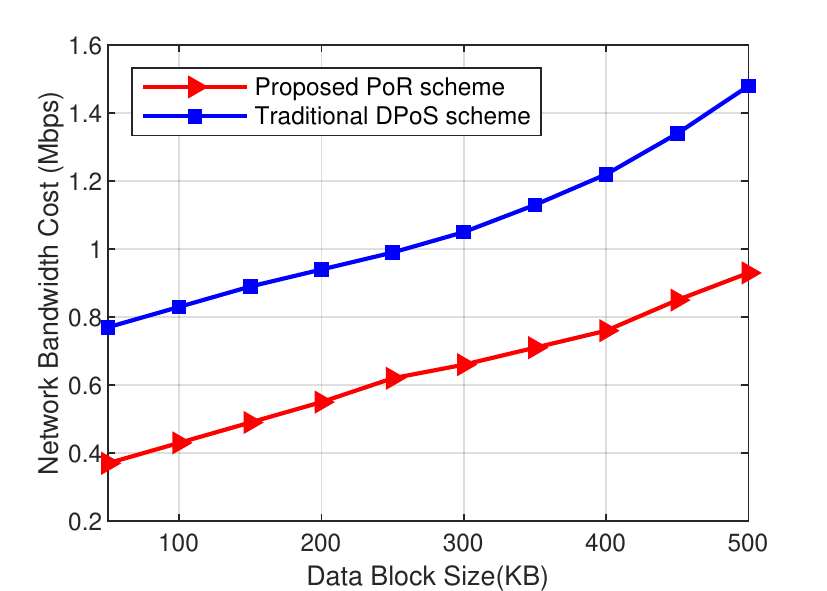} 
			\caption{Comparison of network bandwidth cost.  }
		\end{subfigure}
		\caption{Evaluation of blockchain performance.}
		\label{Blockchain_Performance}
		\vspace{-0.1in}
	\end{figure*}
	
	\begin{figure}
		\centering
		\includegraphics[width=0.99\linewidth]{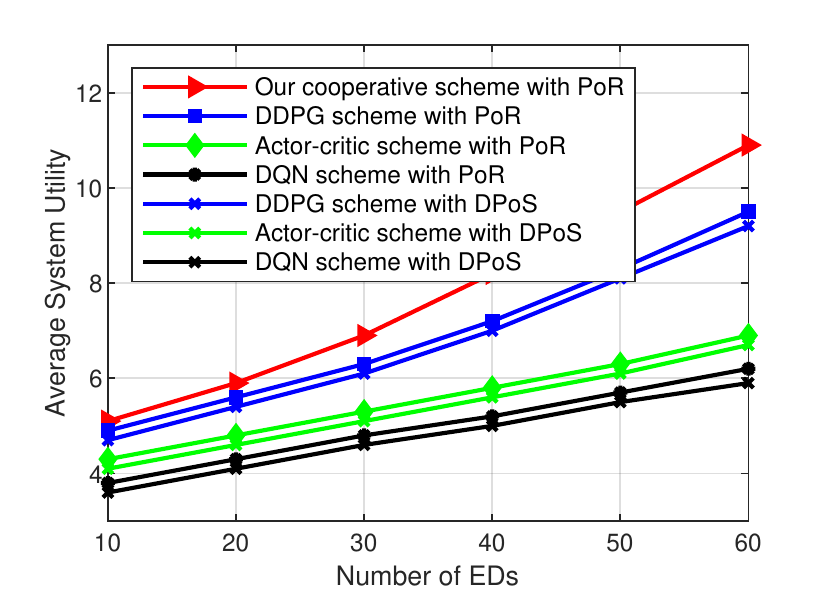} 
		\caption{{Comparison of system utility with non-cooperative schemes. }}
		\label{Fig:SystemUtility1}
		\vspace{-0.1in}
	\end{figure}
	
	\begin{figure}
		\centering
		\includegraphics[width=0.99\linewidth]{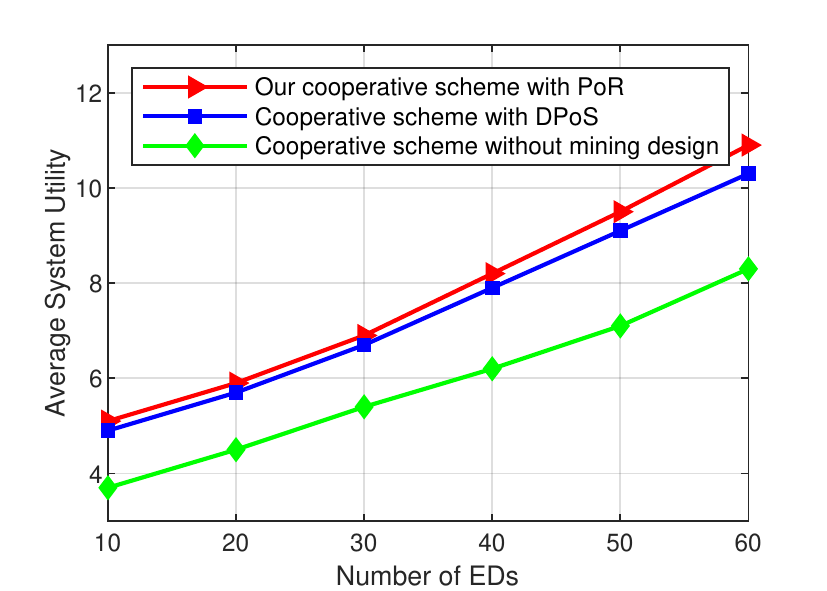} 
		\caption{{Comparison of system utility with cooperative schemes.  }}
		\label{Fig:SystemUtility2}
		\vspace{-0.1in}
	\end{figure}
	We first show the block verification latency performance versus different numbers of mining nodes with the block size fixed at 50 KB and compare with the traditional DPoS scheme \cite{sun2020joint}. As illustrated in Fig.~\ref{Blockchain_Performance}(a), our proposed PoR scheme requires significantly less time for mining blocks, compared to the DPoS scheme thanks to the optimized block verification procedure. Although the time required for block verification increases with the increasing number of miners, our solution still achieves a much better performance than that of the DPoS scheme. This result confirms our lightweight consensus design that is thus well suitable for large-scale blockchain-based MEC systems.
	\begin{figure*}[t!]
		\centering
		\begin{subfigure}[t]{0.5\textwidth}
			\centering
			\includegraphics[width=0.99\linewidth]{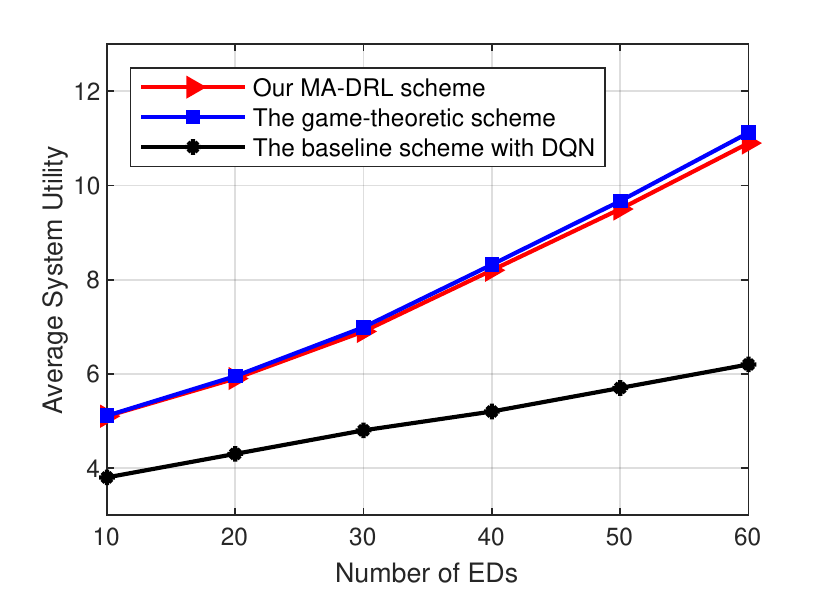} 
			\caption{Average system utility with different numbers of EDs.}
		\end{subfigure}%
		~
		\begin{subfigure}[t]{0.5\textwidth}
			\centering
			\includegraphics[width=0.99\linewidth]{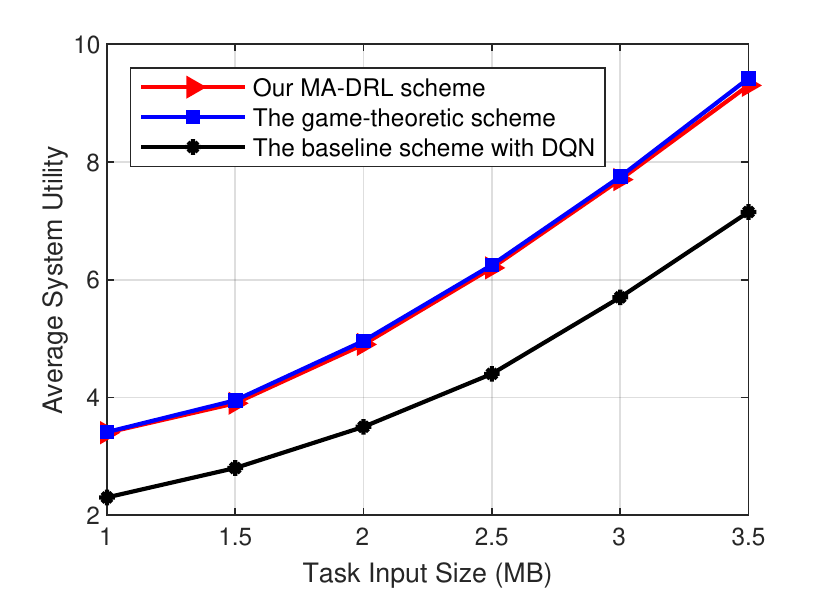} 
			\caption{Average system utility with different IoT data task sizes. }
		\end{subfigure}
		\caption{{\color{black}Comparison of learning and game-theoretic approaches.}}
		\label{Game_Perfocemance}
		\vspace{-0.1in}
	\end{figure*}
	
	Next, Fig.~\ref{Blockchain_Performance}(b) indicates the simulation result in terms of network bandwidth cost spent by the mining process for different data block sizes from 50 KB to 500 KB in the edge blockchain network. Due to the optimized block exchange procedure where each ES only needs to contact with one different miner for transaction verification, instead of using a repeated process, our PoR scheme can save much network bandwidth resources, compared to the DPoS scheme \cite{sun2020joint}. 
	\subsection{Evaluation of Overall System Utility Performance}
	In this subsection, we evaluate the performance of the overall system utility of our TOBM scheme enabled by the joint consideration of offloading utility and mining utility. The performances of our cooperative scheme with our PoR mining design and other non-cooperative schemes with PoR and DPoS mining are illustrated in Fig.~\ref{Fig:SystemUtility1}. Unsurprisingly, our TOBM scheme with a PoR mining design achieves the best overall system utility. The reasons for this observation are two-fold. First, our offloading scheme with a cooperative MA-DDPG algorithm outperforms other non-cooperative offloading schemes in terms of a better offloading utility, as evidenced in  Fig.~\ref{Offloading_Performance}. Second, our PoR design yields a lower mining latency which consequently increases the mining utility, as explained in Section~\ref{Section:MinerNodeSelection}. As a result, our scheme with a cooperative offloading design and a lightweight mining design achieves a much better overall system utility than the other non-cooperative offloading schemes with DPoS design. Moreover, due to its better mining utility, our PoR design contributes to better overall system utilities in each non-cooperative offloading scheme, compared to the use of DPoS design. 
	

	Furthermore, we compare the system utility performance of our cooperative TOBM scheme with other cooperative schemes, including a cooperative scheme with DPoS design \cite{related1} and a cooperative scheme without mining design \cite{21}. As shown in Fig.~\ref{Fig:SystemUtility2}, our TOBM scheme with PoR design achieves a better system utility than the cooperative scheme with DPoS design, thanks to the better mining utility of our PoR framework. Meanwhile, the cooperative scheme in \cite{21} has lowest system utility due to the lack of consideration of mining design.

	{\color{black}\subsection{Comparison of Learning and Game-theoretic Approaches}
		Next, we compare the utility performance between the MA-DRL scheme and the game-theoretic scheme, where the DQN scheme is used as the baseline, after averaging the results from 5 simulations. As shown in Fig.~\ref{Game_Perfocemance}(a), the game-theoretic approach can achieve the optimal system utility, compared to the MA-DRL scheme, when increasing the number of MDs. This is because  in the game approach, each MD can obtain the full knowledge of other MDs’ information such as the information of offloading decisions and mining status via the collaborative interactions. This allows each MD to determine the optimal computation offloading strategy to achieve the converged point of NE. Meanwhile, the MA-DRL scheme can also achieve reasonably close results, where the physical parameters of MDs are time-varying, and each MD can compute the approximately optimal computation offloading strategy without requiring any prior information about other MDs. Similar performances can also be seen in Fig.~\ref{Game_Perfocemance}(b), when increasing the size of task inputs. 
		

		\section{Conclusions and Future Works}
		\label{Conclusion}
		In this article, we have proposed a novel cooperative TOBM scheme to enable a joint design of task offloading and blockchain mining in blockchain-based MEC systems. First, we have proposed a new cooperative  offloading framework that enables EDs to learn offloading policies in a collaborative manner. Then, we have designed a new PoR mining scheme enabled by a lightweight block verification strategy. To this end, we have formulated a joint offloading and mining optimization problem which is solved by an MA-DRL algorithm.  \textcolor{black}{We then derived a game-theoretic solution to model the competition among EDs in offloading and mining as a potential game, and proved the existence of a pure Nash equilibrium.} Simulation results have clearly showed the significant advantages of our proposed scheme over the existing schemes in terms of higher system rewards with better offloading utility and lower blockchain costs which thus enhance the overall system utility.
		
		Our proposed approach has  potential for future intelligent mobile networks, where EDs are able to build distributed intelligent solutions via our cooperative DRL model for enabling intelligent computation, communications and network control \cite{nguyen20216g}. In future work, it is of interest to consider fair resource allocation strategies for simultaneously supporting the edge computation and blockchain services. The tradeoff between mining security and latency should be also studied to strike a beneficial  balance between these two important design factors before integrating into MEC. Moreover, resource trading solutions should be developed to enable reliable energy purchase for resource-constrained edge nodes in blockchain-based MEC systems.
		
		\bibliography{Ref}
		\bibliographystyle{IEEEtran}
		
		\begin{IEEEbiography}[{\includegraphics[width=1in,height=1.25in,clip,keepaspectratio]{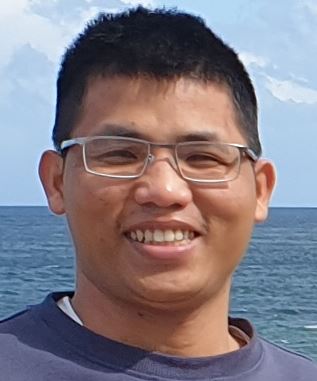}}]{Dinh C. Nguyen}
			(Member, IEEE) is currently working toward the Ph.D. degree with the School of Engineering, Deakin University, Victoria, Australia. His research interests focus on wireless communications, federated learning, deep reinforcement learning, blockchain, and edge computing. He has published over 20 papers as the first author at the top-tier IEEE journals and conferences, such as IEEE Transactions on Mobile Computing, IEEE Wireless Communications Magazine, IEEE Communications Surveys and Tutorials, IEEE Internet of Things Journal, IEEE GLOBECOM, ICC, and CCGrid conferences. He has been a recipient of the prestigious Data61 PhD scholarship, CSIRO, Australia. He has been the TPC member of top-tier conferences including IEEE GLOBECOM 2021.
		\end{IEEEbiography}
		\vskip -2\baselineskip 
		\begin{IEEEbiography}[{\includegraphics[width=1in,height=1.25in,clip,keepaspectratio]{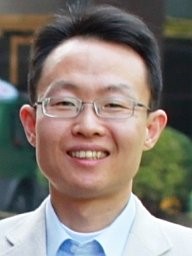}}]{Ming Ding}
			(Senior Member, IEEE) received the B.S. and M.S. degrees (with first-class Hons.) in electronics engineering from Shanghai Jiao Tong University (SJTU), Shanghai, China, and the Doctor of Philosophy (Ph.D.) degree in signal and information processing from SJTU, in 2004, 2007, and 2011, respectively. From April 2007 to September 2014, he worked at Sharp Laboratories of China in Shanghai, China as a Researcher/Senior Researcher/Principal Researcher. Currently, he is a senior research scientist at Data61, CSIRO, in Sydney, NSW, Australia. His research interests include information technology, data privacy and security, machine learning and AI, etc. He has authored over 140 papers in IEEE journals and conferences, all in recognized venues, and around 20 3GPP standardization contributions, as well as a Springer book ``Multi-point Cooperative Communication Systems: Theory and Applications". Also, he holds 21 US patents and co-invented another 100+ patents on 4G/5G technologies in CN, JP, KR, EU, etc. Currently, he is an editor of IEEE Transactions on Wireless Communications and IEEE Wireless Communications Letters. Besides, he has served as Guest Editor/Co-Chair/Co-Tutor/TPC member for many IEEE top-tier journals/conferences and received several awards for his research work and professional services.
		\end{IEEEbiography}
		\vskip -2\baselineskip 
		\begin{IEEEbiography}[{\includegraphics[width=1in,height=1.25in,clip,keepaspectratio]{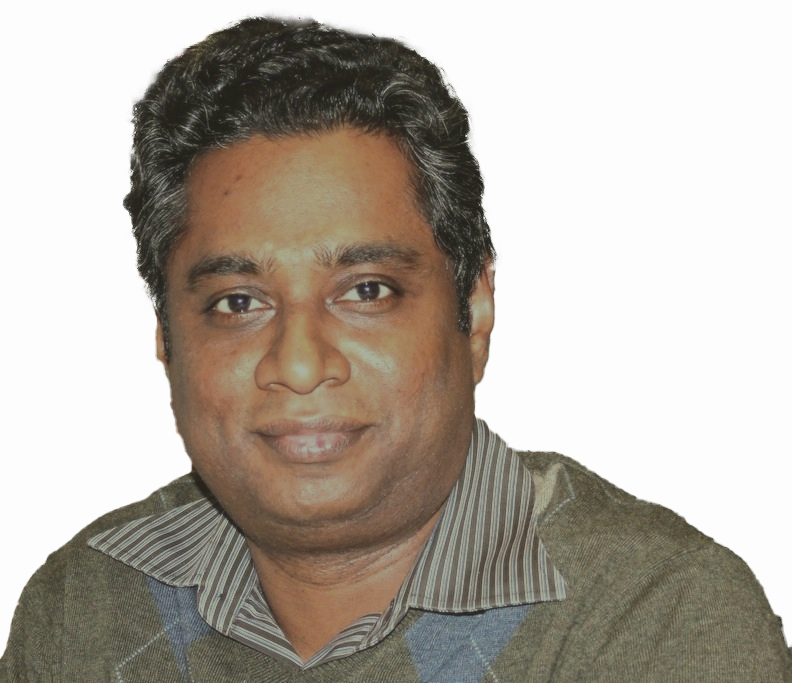}}]{Pubudu N. Pathirana}
			(Senior Member, IEEE) was born in 1970 in Matara, Sri Lanka, and was educated at Royal College Colombo. He received the B.E. degree (first class honors) in electrical engineering and the B.Sc. degree in mathematics in 1996, and the Ph.D. degree in electrical engineering in 2000 from the University of Western Australia, all sponsored by the government of Australia on EMSS and IPRS scholarships, respectively. He was a Postdoctoral Research Fellow at Oxford University, Oxford, a Research Fellow at the School of Electrical Engineering and Telecommunications, University of New South Wales, Sydney, Australia, and a Consultant to the Defence Science and Technology Organization (DSTO), Australia, in 2002. He was a visiting professor at Yale University in 2009. Currently, he is a full Professor and the Head of Discipline, Mechatronics, Electrical and Electronic Engineering and the Director of Network Sensing and Biomedical Engineering(NSBE) research group at the School of Engineering, Deakin University, Geelong, Australia. His current research interests include Bio-Medical assistive device design, human motion capture, mobile/wireless and IoT networks, rehabilitation robotics and signal processing.
		\end{IEEEbiography}
		\vskip -2\baselineskip 
		\begin{IEEEbiography}[{\includegraphics[width=1in,height=1.25in,clip,keepaspectratio]{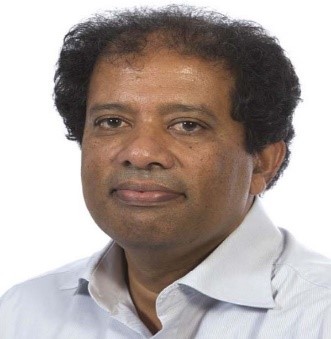}}]{Aruna Seneviratne}
			(Senior Member, IEEE) is currently a Foundation Professor of telecommunications with the University of New South Wales, Australia, where he holds the Mahanakorn Chair of telecommunications. He has also worked at a number of other Universities in Australia, U.K., and France, and industrial organizations, including Muirhead, Standard Telecommunication Labs, Avaya Labs, and Telecom Australia (Telstra). In addition, he has held visiting appointments at INRIA, France. His current research interests are in physical analytics: technologies that enable applications to interact intelligently and securely with their environment in real time. Most recently, his team has been working on using these technologies in behavioral biometrics, optimizing the performance of wearables, and the IoT system verification. He has been awarded a number of fellowships, including one at British Telecom and one at Telecom Australia Research Labs.
		\end{IEEEbiography}
		\vskip -2\baselineskip 
		\begin{IEEEbiography}[{\includegraphics[width=1in,height=1.25in,clip,keepaspectratio]{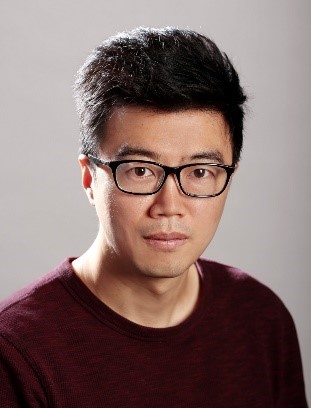}}]{Jun Li}
			(M'09-SM'16) received Ph. D degree in Electronic Engineering from Shanghai Jiao Tong University, Shanghai, P. R. China in 2009. From January 2009 to June 2009, he worked in the Department of Research and Innovation, Alcatel Lucent Shanghai Bell as a Research Scientist. From June 2009 to April 2012, he was a Postdoctoral Fellow at the School of Electrical Engineering and Telecommunications, the University of New South Wales, Australia. From April 2012 to June 2015, he is a Research Fellow at the School of Electrical Engineering, the University of Sydney, Australia. From June 2015 to now, he is a Professor at the School of Electronic and Optical Engineering, Nanjing University of Science and Technology, Nanjing, China. He was a visiting professor at Princeton University from 2018 to 2019. His research interests include network information theory, game theory, distributed intelligence, multiple agent reinforcement learning, and their applications in ultra-dense wireless networks, mobile edge computing, network privacy and security, and industrial Internet of things. He has co-authored more than 200 papers in IEEE journals and conferences, and holds 1 US patents and more than 10 Chinese patents in these areas. He was serving as an editor of IEEE Communication Letters and TPC member for several flagship IEEE conferences. He received Exemplary Reviewer of IEEE Transactions on Communications in 2018, and best paper award from IEEE International Conference on 5G for Future Wireless Networks in 2017.
		\end{IEEEbiography}
		\vskip -2\baselineskip 
		\begin{IEEEbiography}[{\includegraphics[width=1in,height=1.25in,clip,keepaspectratio]{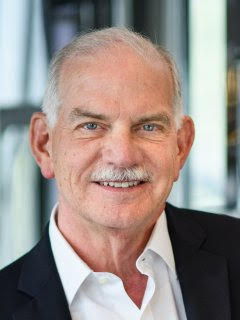}}]{H. Vincent Poor}
			(S'72, M'77, SM'82, F'87) received the Ph.D. degree in EECS from Princeton University in 1977. From 1977 until 1990, he was on the faculty of the University of Illinois at Urbana-Champaign. Since 1990 he has been on the faculty at Princeton, where he is currently the Michael Henry Strater University Professor. During 2006 to 2016, he served as the dean of Princeton’s School of Engineering and Applied Science. He has also held visiting appointments at several other universities, including most recently at Berkeley and Cambridge. His research interests are in the areas of information theory, machine learning and network science, and their applications in wireless networks, energy systems and related fields. Among his publications in these areas is the forthcoming book \textit{Machine Learning and Wireless Communications}, (Cambridge University Press). Dr. Poor is a member of the National Academy of Engineering and the National Academy of Sciences and is a foreign member of the Chinese Academy of Sciences, the Royal Society, and other national and international academies. He received the IEEE Alexander Graham Bell Medal in 2017.
		\end{IEEEbiography}
	\end{document}